\documentclass[aps,prb,twocolumn,notitlepage]{revtex4-2}

\usepackage{amsmath}
\usepackage{amssymb}
\usepackage{graphicx}
\usepackage{epstopdf}
\usepackage{bm}
\usepackage[colorlinks=true]{hyperref}

\begin{document}

\title{Superfluid-like spin transport in the dynamic states of easy-axis magnets}

\author{Jongpil Yun}
\author{Se Kwon Kim}
\affiliation{Department of Physics, Korea Advaced Institute of Science and Technology, Daejeon 34141, Republic of Korea}

\date{\today}

\begin{abstract}
The existing proposals for superfluid-like spin transport have been based on easy-plane magnets where the U(1) spin-rotational symmetry is spontaneously broken in equilibrium, and this has been limiting material choices for realizing superfluid-like spin transport to restricted class of magnets. In this work, we lift this limitation by showing that superfluid-like spin transport can also be realized based on easy-axis magnets, where the U(1) spin-rotational symmetry is intact in equilibrium but can be broken in non-equilibrium. Specifically, we find the condition to engender a non-equilibrium easy-cone state by applying a spin torque to easy-axis magnets, which dynamically induces the spontaneous breaking of the U(1) spin-rotational symmetry and thereby can support superfluid-like spin transport. By exploiting this dynamic easy-cone state, we show theoretically that superfluid-like spin transport can be achieved in easy-axis magnets under suitable conditions and confirmed the prediction by micromagnetic simulations. We envision that our work broadens material library for realizing superfluid-like spin transport, showing the potential utility of dynamic states of magnets as venue to look for spin-transport phenomena that do not occur in static magnetic backgrounds.
\end{abstract}

\maketitle
\section{Introduction}
Spintronics is a field that harnesses spin degree of freedom of electrons to store, transport, and process information in order to go beyond the conventional electronics where only charge degree of freedom has been used~\cite{Igorspintronics2004, Wolfspintronics2001, Guorecent2019}. Since information is encoded in the form of spin, it is important to achieve efficient spin transport in spintronics, which demands us to identify and employ low-dissipation magnetic materials and spin transport therein. In this regard, magnetic insulators have emerged as energy-wise efficient material platforms for spintronics since they have no Joule heating associated with an electric current and therefore can enable low-power spin-based information transport and processing. In a magnetic insulator, spin is transported by its collective excitations, i.e., spin waves, whose quanta are called magnons~\cite{Kittel, Kajiwaratransmission2010, Uchidaspin2010, Chumakmagnon2015}. Since magnon-based information transport and processing can be realized without the Joule heating in principle, generating and controlling magnons have been actively investigated to realize magnonic devices~\cite{Navabicontrol2019, Barsukovgiant2019, Etesamiradcontrol2021}, which includes the experimental demonstration of long-distance spin transport in certain magnets~\cite{Cornelissenlong2015, Cornelissenmagnon2016, Wesenberglong2017}. Most of the previously studied magnonic spin transport have been based on diffusion of magnons, which has a critical problem in that the spin current exponentially decays away from the spin-current source~\cite{Cornelissenlong2015}. To circumvent this problem of rapidly decaying spin current of diffusive magnons, a novel type of spin transport referred to as superfluid-like spin transport has emerged as an alternative for efficient spin transport~\cite{Soninspin2010}.

Superfluid-like spin transport is a spin-analogue of mass superfluidity. The mass superfluidity can occur when the wave function defined by $\psi=\sqrt{n}e^{i\theta}$, where $n$ is a particle density and $\theta$ is an arbitrary phase, breaks the U(1) phase symmetry spontaneously~\cite{Oxfordsuper2004}. Likewise, superfluid-like spin transport can occur in magnetically ordered systems when the magnetic order parameter breaks the U(1) spin-rotational symmetry spontaneously~\cite{Soninspin2010, Soninspin2013}. In contrast to the exponential decaying of diffusive spin transport, superfluid-like spin transport has the characteristic of algebraically decaying spin current, which enables long-distance spin transport in certain magnets~\cite{Konigdissipationless2001, Chennonlocal2014, Chenspin2014, Hansspin2015, Chendissi2015, Lindersuper2015, Flebustwofluid2016, Kiminteraction2016, Chengtera2016, Qaiumzadehspin2017}. The existing research of superfluid-like spin transport has been focused only on easy-plane magnets in which the system breaks the U(1) spin-rotational symmetry spontaneously in equilibrium~\cite{Takeiafm2014, Takeifm2014, Takeinonlocal2015, Soninsuperfluid2019}.

\begin{figure}
\includegraphics[width=\linewidth]{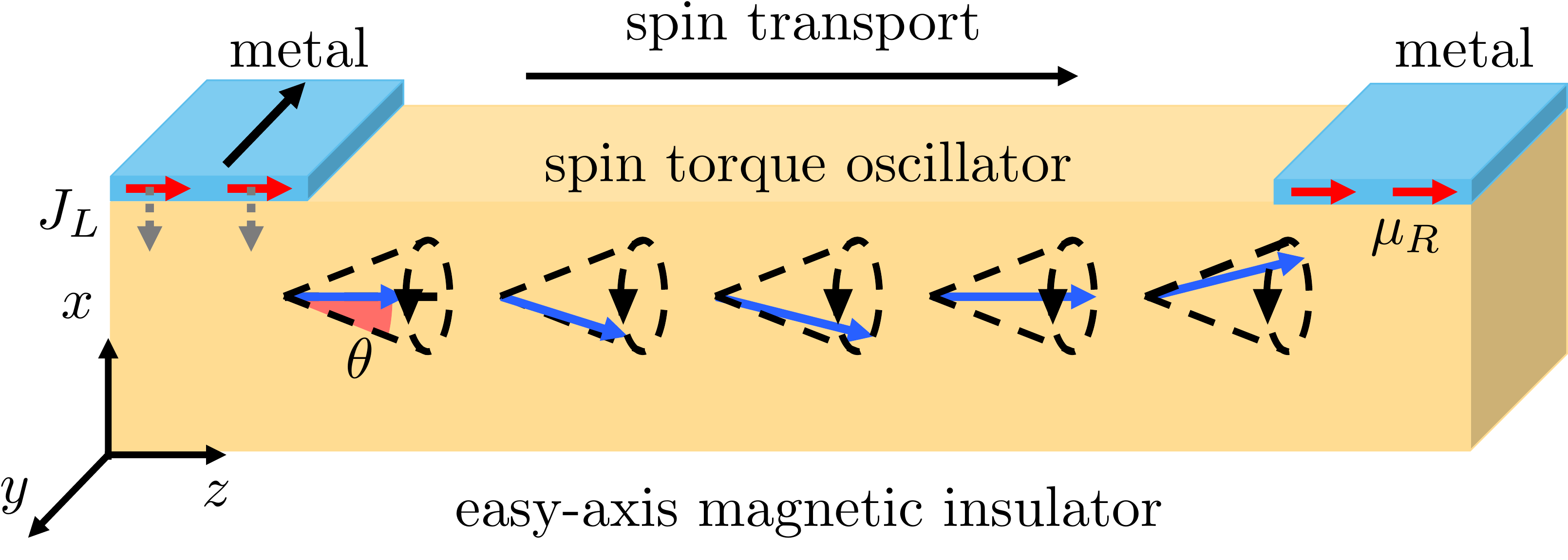}
\caption{Schematics of the experimental setup for realizing superfluid-like spin transport in a magnetic insulator having the $z$ axis as an easy axis subjected to a spin torque, denoted by a spin-torque oscillator. In the magnetic insulator, the blue arrows depict the spatially varying magnetization and the dashed cones represent the precession trajectories (forming cones with tilting angle $\theta$ from the easy axis) of the local spins driven by the spin torque. The black arrow of the left metal represents the charge current, which injects the spin current $\left(J_L\right)$ to the left end of the ferromagnet via the spin Hall effect. The red arrows of the left and the right metals are the direction of spin accumulations, at the interfaces between the metals and the ferromagnet. The spin accumulation $\mu_R$ at the right metal generated by the nonlocal spin transport from the left metal can be detected via the inverse spin Hall effect.}
\label{fig:fig1}
\end{figure}

In this work, for a potential material platform for superfluid-like spin transport, we consider easy-axis magnets, where the magnetization aligns with easy-axis and thus does not break U(1) spin-rotational symmetry spontaneously in equilibrium. Instead of using the equilibrium U(1) symmetry breaking as done for previous proposals based on easy-plane magnets, we turn to the dynamic breaking of the U(1) spin-rotational symmetry. More specifically, to break U(1) spin-rotational symmetry spontaneously in easy-axis magnets, we apply a spin torque and engender a dynamic easy-cone state to support superfluid-like spin transport~\cite{slonczewskicurrent1996, Ralphspin2008, Yaroslavtheory2008, Xingtaospin2011, Duannanowire2014, Yaroslavpers2018, Flebusnon2020}. The system is schematically illustrated in Fig.~\ref{fig:fig1}. The easy-axis magnet subjected to a suitable spin torque forms a spin-torque oscillator, in which the local magnetization (shown as the blue arrows) precesses within cones (depicted by the dashed black lines with the cone angle $\theta$) by breaking the U(1) spin-rotational symmetry dynamically. Applying a charge current in the left metal injects a spin current $J_\text{L}$ from the left metal to the left end of the magnet via the spin Hall effect~\cite{Katoobser2004, Xingtaospin2011, Yaroslavspin2014, Sinovaspin2015}. The injected spin current is transported through the magnet by superfluid-like spin transport in the form of the spatially varying order parameter. The spin transport generates a spin accumulation $\mu_\text{R}$ at the interface between the ferromagnet and the right metal, which can be probed either by spin pumping~\cite{Yaroslavenhanced2002, Yaroslavnonlocal2005} or by the inverse spin Hall effect. In our system, non-local spin transport refers to the generation of the spin accumulation $\mu_\text{R}$ at the right end of the ferromagnet by the spin-current injection from the left end. By the theoretical analysis based on the Landau-Lifshitz-Gilbert (LLG) equation~\cite{Quantum, GilbertIEEE2004} and the micromagnetic simulations, we show that the spin accumulation $\mu_\text{R}$ decays algebraically as the ferromagnet length increases, exhibiting superfluid-like spin transport.

The paper is organized as follows. In Sec.~\ref{sec:model}, we describe the model system, namely the easy-axis ferromagnet subjected to a spin torque, and identify the condition with which superfluid-like spin transport can be realized by theoretical analysis and micromagnetic simulations. In Sec.~\ref{sec:transport}, we theoretically and numerically show that our system can indeed support algebraically decaying spin current, i.e., superfluid-like spin transport. In Sec.~\ref{sec:summary}, we summarize our results.

\section{Model}
\label{sec:model}

Our model system which is illustrated in Fig.~\ref{fig:fig1} is a quasi-one-dimensional ferromagnetic wire whose energy is given by
\begin{equation}
U=\int dV \left[\frac{A\mathbf{m}'^2-K_\text{eff} m_z^2+K_2 m_z^4}{2}-\mathbf{H} \cdot\mathbf{m}\right],
\label{eq:U}
\end{equation}
where $\mathbf{m}$ is the three-dimensional unit vector in the direction of the magnetization, the prime ($'$) is the gradient with respect to the $z$-coordinate along the wire, $A$ is the exchange coefficient, $K_\text{eff}>0$ is the first-order effective anisotropy which combines the shape anisotropy and the first-order easy-axis crystalline anisotropy, $K_2 > 0$ is the second-order anisotropy~\cite{Riespin2015, Shawperp2015, Jangspin2016, Jangdomain2019, Gweonintrinsic2020, Gweonevolution2020}, and $\mathbf{H}$ is the external magnetic field. We assume that the system is quasi-one-dimensional so that the magnetization varies only along the $z$ direction: $\mathbf{m}(z, t)$. The external magnetic field is applied along the easy-axis direction: $\mathbf{H}=H\hat{\mathbf{z}}$. We consider the cases where the ground state is given by the uniform magnetization in the $z$ direction $\mathbf{m}(z, t) \equiv\hat{\mathbf{z}}$, which is satisfied when $K_2 < \left(K_\text{eff}+H\right)/2$. Note that the energy $U$ is invariant under uniform rotations of the magnetization about the $z$-axis, i.e., $\mathbf{m} \mapsto R_z \mathbf{m}$ with an arbitrary rotation matrix $R_z$ about the $z$ axis, indicating that the system possess the U(1) spin-rotational symmetry about the $z$ axis. Since the ground state $\mathbf{m}(z, t) \equiv \hat{\mathbf{z}}$ is invariant under the spin rotations, it does not break the U(1) spin-rotational symmetry.

The equation of motion for the dynamics of the magnetization $\mathbf{m}$ subjected to a spin torque is given by the LLG equation~\cite{Quantum, GilbertIEEE2004} augmented by the spin-torque term :
\begin{equation}
s\dot{\mathbf{m}}-\alpha s \mathbf{m}\times \dot{\mathbf{m}}=-\mathbf{m}\times\mathbf{h}_\text{eff}+\bm{\tau}_\text{ST},\label{llg}
\end{equation}
where $s$ is the saturated (scalar) spin density, the dot $\left(\dot{}\right)$ denotes differentiation with respect to time, $\alpha>0$ is the dimensionless Gilbert damping parameter, $\mathbf{h}_\text{eff}=-\delta U/\delta\mathbf{m}$ is the effective field, and $\bm{\tau}_\text{ST}=\tau_\text{ST}\mathbf{m}\times\left(\mathbf{m}\times\hat{\mathbf{z}}\right)$ is a externally-applied spin torque polarized along the $z$ direction, which can be realized either by the spin-transfer torque~\cite{Ralphstt} or by spin-orbit torque~\cite{Manchonsot}. We assume that this spin torque $\bm{\tau}_\text{ST}$ is exerted uniformly on the ferromagnet. 

To endow the ferromagnet with the capability to support superfluid-like spin transport, it is necessary to induce the ferromagnet to break the U(1) spin-rotational symmetry dynamically, which can be done by driving it into a self-oscillatory mode with the sufficiently strong spin torque. The detailed condition for this oscillating phase can be obtained as follows. The LLG equation in terms of the polar angle $\left(\theta\right)$ and the azimutal angle $\left(\phi\right)$ with $\mathbf{m}=\sin\theta\cos\phi\hat{\mathbf{x}}+\sin\theta\sin\phi\hat{\mathbf{y}}+\cos\theta\hat{\mathbf{z}}$ is given by
\begin{align}
s\left(\dot{\theta}\sin{\theta}+\alpha\dot{\phi}\sin^2{\theta}\right)=&A\left(\phi'\sin^2{\theta}\right)'+\tau_\text{ST}\sin^2{\theta},\label{fmllg1}
\\
\begin{split}
s\left(\dot{\phi}\sin{\theta}-\alpha\dot{\theta}\right)=&A\left(\phi'^2\sin{\theta} \cos{\theta}-\theta''\right) \\
&+ H \sin{\theta}+K_\text{eff}\sin{\theta} \cos{\theta}\\
&-2K_2\sin\theta\cos^3\theta.
\end{split}\label{fmllg2}
\end{align}
Equation~\eqref{fmllg1} has a clear physical meaning: It is the spin continuity equation: The first term and the second term on the left-hand side are the time evolution of the $z$ component of spin density and the damping term, respectively. The first term on the right-hand side of Eq.~\eqref{fmllg1} is the divergence of spin current density: $j^s=-A\sin^2\theta \, \partial_x \phi$ and the second term is the spin current coming from the bulk spin torque.

Now, we look for a condition under which the spin torque induces a dynamic easy-cone state and thus the spontaneous breaking of the U(1) spin-rotational symmetry. A steady-state solution of Eqs.~\eqref{fmllg1} and~\eqref{fmllg2} with constant polar angle with $\dot{\theta} = 0$ satisfies
\begin{align}
\tau_\text{ST}&=\alpha\left(K_\text{eff}\cos\theta+H-2K_2\cos^3\theta\right) \, . \label{ttheta}
\end{align}
Then, the condition that the system is in a dynamic easy-cone state with $0 < \theta < \pi$ is given by
\begin{align}
\alpha (K_\text{eff} + H - 2 K_2) <&\tau_\text{ST}< \alpha\left(\frac{2}{3}\sqrt{\frac{K_\text{eff}}{6K_2}}K_\text{eff}+H\right) \, ,\label{taucon}\\
\frac{K_\text{eff}}{6}< &K_2<\frac{K_\text{eff}+H}{2} \, , \label{sanis}
\end{align}
where $\tau_{c,1} = \alpha (K_\text{eff} + H - 2 K_2)$ is the lower critical torque that is given by the value of the right-hand side of Eq.~(\ref{ttheta}) at $\theta = 0$ and $\tau_{c,2} = \alpha\left(\frac{2}{3}\sqrt{\frac{K_\text{eff}}{6K_2}}K_\text{eff}+H\right)$ is the upper critical torque. The lower critical torque $\tau_{c,1}$ is the minimum torque that is required to drive the ferromagnet into the dynamic easy-cone state, i..e, self-oscillatory state. When we apply the torque lower than the lower critical torque, the magnetization is kept along the $z$ axis without breaking the U(1) spin-rotational symmetry. The physical meaning of the upper critical field is as follows. When we apply the spin torque larger than the upper critical torque $\tau_{c,2}$, the magnetization reverses into the negative $z$ direction completely, i.e, $\mathbf{m} = - \hat{\mathbf{z}}$, which does not show the dynamic oscillation of the magnetization. Equation~\eqref{sanis} states the condition for the system parameters, $K_\text{eff}, K_2$, and $H$. When $K_2$ is smaller than $K_\text{eff}/6$, the system does not possess a dynamic easy-cone state, but is saturated along either the positive $z$ direction (for weak spin torques) or the negative $z$ direction (for strong spin torques). The condition $K_2<(K_\text{eff}+H)/2$ is to ensure that the unperturbed ground state of the ferromagnet is given by the $z$ direction, as discussed above. When the spin torque and the system parameters satisfy Eqs.~\eqref{taucon} and~\eqref{sanis}, the system is driven into a dynamic easy-cone state, where the magnetization precesses about the $z$ axis with arbitrary initial values for the azimuthal angle $\phi$ and thereby spontaneously break the U(1) spin-rotational symmetry.

\begin{figure}[htp!]
\includegraphics[width=0.7\linewidth]{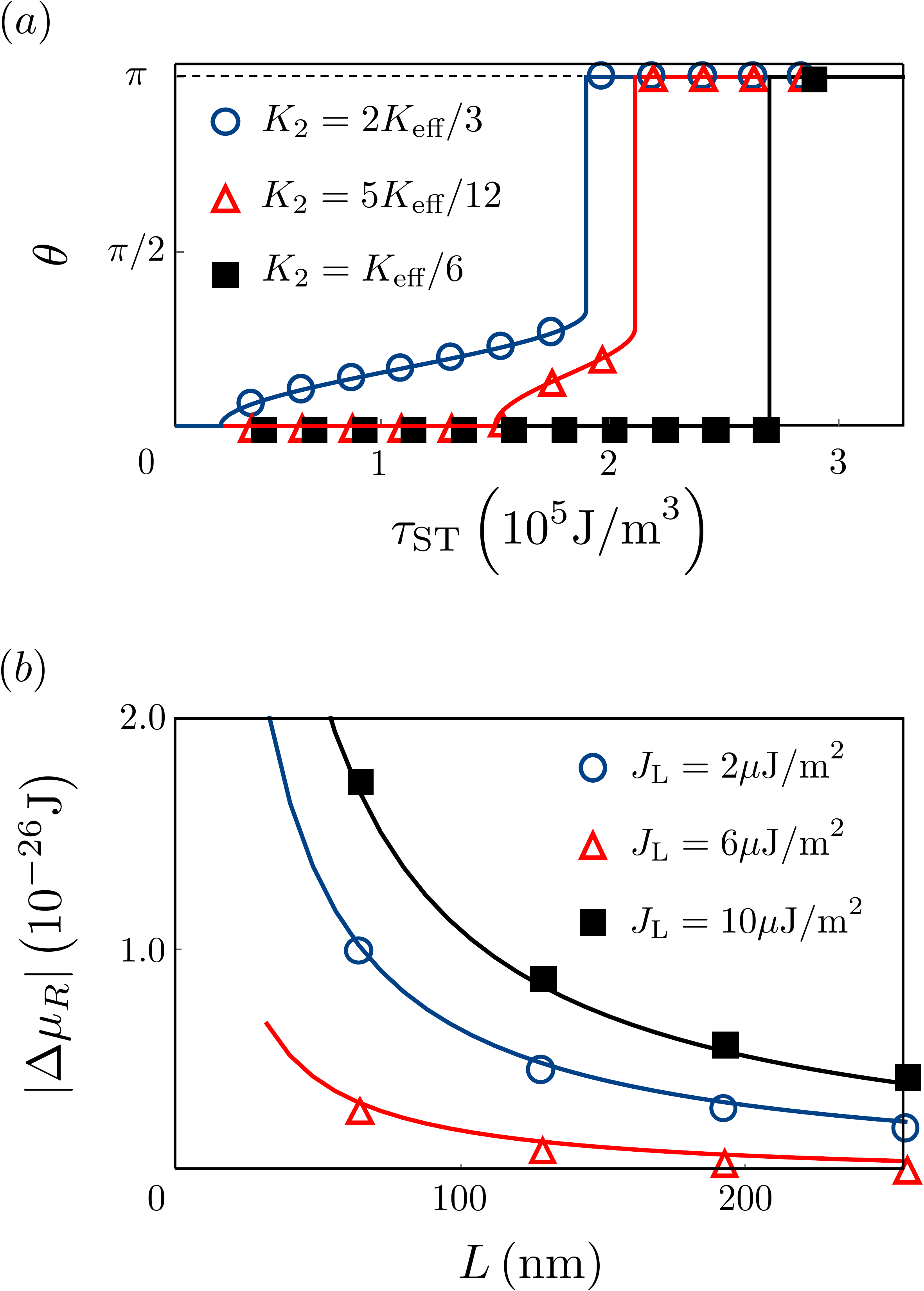}
\caption{(a) The polar angle $\left(\theta\right)$ of the magnetization as a function of the bulk spin torque $\left(\tau_\text{ST}\right)$. The lines show the theoretical results [Eq.~\eqref{ttheta}] and the symbols represent the simulation results. The circles, the triangles, and the squares correspond to $K_2=2K_\text{eff}/3$, $K_2=5K_\text{eff}/12$, and $K_2=K_\text{eff}/6$, respectively. (b) The spin accumulation $\left(\lvert\Delta\mu_R\rvert\right)$ at the right boundary of the ferromagnet induced by a spin-current injection $J_L$ from the left boundary as a function of the ferromagnet length $\left(L\right)$. The lines show the theoretical result [Eq.~\eqref{fmaccum}] and the symbols correspond to the simulations results. The circles, the triangles, and the squares correspond to input spin current $J_L = 2\times10^{-6}$J/m$^2$, 6$\times10^{-6}$J/m$^2$ and 10$\times10^{-6}$J/m$^2$, respectively.}
\label{fig:fig2}
\end{figure}

To confirm the spontaneous U(1) spin-rotatinal symmetry breaking under conditions [Eqs.~\eqref{taucon} and~\eqref{sanis}], we run the micromagnetic simulation using $\textsc{MuMax}^3$~\cite{Vansmumax32014} with the following material parameters of NdCo$_5$: $K_\text{eff}=2.4\times10^6\text{J/m}^3$, $H=1\text{T},$ $M_s=1.1\times10^6\text{A/m},$ $A=1.05\times10^{-11}\text{J/m}$ and $\alpha=0.1$~\cite{Ohkoshispin1976, Seifertdomain2013, Mariettamicro2017, Santoshtorque2020}. The demagnetization effects are captured by the first-order effective anisotropy $K_\text{eff}$ through the shape anisotropy in our simulations, as done analytically in Refs.~\cite{Shawperp2015, TimopheevSR2016, Gweonevolution2020, Gweonintrinsic2020}, which is expected to be a good approximation when the system is quasi-one-dimensional cylindrical wire~\cite{Coey, WieserPRB2010, YanPRL2012}. We run the simulation with three different values of second-order anisotropy: $K_2=0.4\times10^6\text{J/m}^3$, $1\times10^6\text{J/m}^3$ and $1.6\times10^6\text{J/m}^3$. For the polar angle in the presence of a spin torque, the analytical result (which can be obtained by solving Eq.~\eqref{ttheta} for $\theta$) and the simulation results are shown in Fig.~\ref{fig:fig2}(a) as the lines and the symbols, respectively, which agree with each other. In Fig.~\ref{fig:fig2}(a), the blue and the red symbols correspond to the cases with $K_2=1.6\times10^6\text{J/m}^3$ and $K_2=1\times10^6\text{J/m}^3$, respectively. For these cases, the ferromagnet is in a steady state with nontrivial polar angle $\theta \neq 0, \pi$, i.e., in a dynamic easy-cone state under suitable spin-torque values. The black symbols correspond to the cases with $K_2 = K_\text{eff} / 6$, where the polar angle is either $0$ or $\pi$ regardless of the spin-torque values as discussed above and the dynamic easy-cone state is not available. 

The obtained dynamic easy-cone state can be interpreted in the framework of the Gross-Pitaevskii equation~\cite{GrossINC1961, PitaevskiiJETP1961} as follows. The LLG equation~\eqref{llg} for the magnetization $\mathbf{m}$ can be recast into the equation for the complex order parameter defined by $\psi (x, t) = m_x (x, t) - i m_y (x, t) = \sqrt{\rho}e^{i\phi}$, where $\rho = \sin \theta$ and $\phi$ are analogous to the density and the phase of the condensate~\cite{Benderdynamic2014, Flebustwofluid2016}. The LLG equation in terms of the complex order parameter $\psi$ is given by
\begin{align}
\begin{split}
is\frac{\partial\psi}{\partial t}=&\left(\left(K_\text{eff}-2K_2+H\right)+\left(-\frac12K_\text{eff}+3K_2\right)\lvert\psi\rvert^2\right)\psi\\
&+s\alpha\left(1-\frac12\lvert\psi\rvert^2\right)\frac{\partial\psi}{\partial t}+i\tau_\text{ST}\left(1-\frac12\lvert\psi\rvert^2\right)\psi \, ,
\end{split}\label{gpllg}
\end{align}
up to the third order in $\psi$. The first term on the right-hand side of the equation originates from the potential energy of our system [Eq.~(\ref{eq:U})], in which $(K_\text{eff}-2K_2+H)$ can be interpreted as the single particle potential and $\left(- K_\text{eff}/2+3K_2\right)$ can be regarded as the interaction strength. The condition to possess a vacuum ground state and the condition to have a stable condensate under pumping (i.e., the repulsive interaction) are respectively given by $K_2<(K_\text{eff}+H)/2$ and $K_\text{eff}/6<K_2$, which are identical to the conditions in Eq.~\eqref{sanis} that were obtained directly from the LLG equation. 

There are two additional terms on the right-hand side in the equation that break the time-reversal symmetry: The second term is the damping term and the third term represents the particle pumping by the spin torque. To have a finite condensate $\psi \neq 0$ in a steady state, the pumping $\propto \tau_\text{ST}$ should be sufficiently strong to dominate the damping term. This condition is given by $\tau_\text{ST} > \alpha (K_\text{eff} + H - 2 K_2)$, which is identical to the lower critical torque that we obtained above. The upper critical torque is not available in Eq.~(\ref{gpllg}), since it is truncated to the third order in the order parameter and thus cannot capture the dynamics of the dense condensate.

\section{superfluid-like spin transport}
\label{sec:transport}

Now, let us investigate the nonlocal spin transport behavior of an obtained dynamic easy-cone state of the easy-axis ferromagnet by assuming that our system satisfies Eqs.~\eqref{taucon} and~\eqref{sanis} so that it breaks the U(1) spin-rotational symmetry dynamically. The situation that we consider is depicted in Fig.~\ref{fig:fig1}. To inject a spin current $J_L$ to the ferromagnet through the left end, one heavy metal with a finite charge current is attached to the left end of the ferromagnet. To detect a spin accumulation $\mu_\text{R}$ at the right end of the ferromagnet, the other heavy metal with no external current is attached to the right end of the ferromagnet.
When we attach the metals to the left and right boundaries of the ferromagnet, there arises two effects: the spin-current injection from the metal with a finite current and spin pumping from the ferromagnet to the metals~\cite{Kiminteraction2016}, which determines the boundary conditions for the spin current at the left end $x=0$ and the right end $x=L$:
\begin{align}
j^s(0)&=J_L\sin^2\theta -\gamma\sin^2\theta \, \dot{\phi}(0),\label{lbound}\\
j^s(L)&=\gamma\sin^2\theta \, \dot{\phi}(L),\label{rbound}
\end{align}
where $j^s = - A \sin^2 \theta \, \partial_x \phi$ is the spin current of the ferromagnet, $L$ is the length of the ferromagnet, $\theta$ is the polar angle of the magnetization, $\gamma = \hbar g^{\uparrow \downarrow} / 4 \pi$, and $g^{\uparrow \downarrow}$ is the effective interfacial spin-mixing conductance between the ferromagnet and the normal metal~\cite{Brataasfinite2000, Yaroslavspin2014}. The first term $\propto J_L$ on the right-hand side of Eq.~\eqref{lbound} is the spin current injected from the left metal to the ferromagnet by the spin Hall effect, where $J_L$ is proportional to the product of the charge current flowing in the left metal and the effective spin Hall angle of the interface between the ferromagnet and the left metal. The second term $\propto \gamma$ on the right-hand side is the spin current ejected from the ferromagnet to the left metal by the spin pumping. The right-hand side of Eq.~\eqref{rbound} is the spin current ejected from the ferromagnet to the right metal by the spin pumping.

By solving the bulk LLG Eq.~\eqref{fmllg1} with the boundary conditions [Eqs.~\eqref{lbound} and~\eqref{rbound}] for a steady state, we obtain the following spin current density and the precessional velocity of the azimuthal angle: 
\begin{align}
\begin{split}
j^s(x,t)&=\left(J_L-(\gamma+\alpha sx)\omega+\tau_\text{ST}x\right)\sin^2\theta ,
\end{split}\label{fmcurrent}\\
\dot{\phi}(x,t)&\equiv\omega=\frac{J_L+\tau_\text{ST} L}{2\gamma+\alpha s L},\label{fmphi}
\end{align}
where the value of the polar angle $\theta$ changes from the value obtained from Eq.~\eqref{ttheta} due to the additional input spin current from the left boundary~\footnote{The modified value of the polar angle $\theta$ can be calculated by solving Eq.~\eqref{fmllg2} and Eq.~\eqref{fmphi} self-consistently for $\theta$.}. The precession of the magnetization induces a finite spin accumulation given by $\mathbf{\mu}_R=-\hbar \hat{\mathbf{z}} \cdot \mathbf{m}\times\dot{\mathbf{m}} = - \hbar \sin^2 \theta \, \dot{\phi}$ with $\hbar$ the reduced Planck constant, which can be measured experimentally~\cite{Hansspin2015, Yaroslavenhanced2002, Yaroslavnonlocal2005}.

To investigate the non-local spin transport from the left end $x = 0$ to the right end $x = L$ through the dynamic ferromagnet, we employ the spin accumulation $\mu_R$ at the interface between the ferromagnet and the right heavy metal and extract the component that is induced by the spin-current injection $J_L$ from the left metal. In other words, we use the difference of the spin accumulation $\mu_R$ between the two cases: with spin-current injection from the left end ($J_L\neq0$) and without the spin-current injection ($J_L=0$). Using $\dot{\phi}$ of Eq.~\eqref{fmphi}, $\Delta\mu_R=\mu_R\left(J_L\neq0\right)-\mu_R\left(J_L=0\right)$ is given by
\begin{equation}
\begin{split}
\Delta\mu_R=&-\left(\frac{J_L}{2\gamma+\alpha s L}\sin^2\theta\right)\hbar \\
&-\frac{\tau_\text{ST} L}{2\gamma+\alpha s L} \left(\sin^2\theta-\sin^2\theta_0\right)\hbar \, ,
\end{split}
\label{fmaccum}
\end{equation}
where $\theta_0$ is the polar angle obtained from Eq.~\eqref{ttheta} in the absence of an input spin current $J_L = 0$. The first term on the right-hand side is the spin accumulation at the right end induced by injecting a spin current $J_L$ at the left end. It decays algebraically $\sim 1/L$ for sufficiently long samples as a function of the ferromagnet length $L$, which is the characteristic superfluid-like spin transport. The second term can be interpreted as the effect of the polar-angle change (from $\theta_0$ to $\theta$) induced by the spin-current injection $J_L$ from the left end. The algebraic decaying behavior of the second term is not evident from the analytical expression above, but we can show that, by linearizing Eq.~(\ref{fmllg2}) with respect to the injected spin current $J_L$, the second term is approximately given by $[2 \hbar s \cos \theta_0 / (6 K_2 \cos^2 \theta_0 - K_\text{eff})] J_L \tau_\text{ST} L / (2 \gamma + \alpha s L)^2$, which decays as $1/L$ for sufficiently long samples. Therefore, the spin accumulation $\Delta \mu_R$ at the right end induced by the spin-current injection from the left end decays as $1/L$ as the ferromagnet length $L$ increases, exhibiting superfluid-like spin transport.

To confirm our theoretical prediction of superfluid-like spin transport in a dynamic cone state of ferromagnets, we perform the micromagnetic simulations and compare the simulation results against the theoretical results [Eq.~\eqref{fmaccum}]. In simulations, we use the same material parameters that were mentioned above and the fixed second-order anisotropy $K_2=1.6\times10^6\text{J/m}^3$. For simplicity, we assumed that the spin pumping effect is negligible by setting $\gamma=0\text{J/m}^2$. Figure~\ref{fig:fig2}(b) plots the spin accumulation difference  $\Delta \mu_R$ between $J_L\neq0$ and $J_L=0$ as a function of the ferromagnet length $L$ for several different values of the input spin current $J_L$. The non-local spin transport $\Delta\mu_R$ decays algebraically, not exponentially, as the ferromagnet length $L$ increases. Our analytical [Eq.~\eqref{fmaccum}] and simulation results [Fig.~\ref{fig:fig2}(b)] show that we can realize superfluid-like spin transport using dynamic states of easy-axis ferromagnets. These are our main results.

\section{Summary}
\label{sec:summary}

To go beyond the previous works on superfluid-like spin transport that have been restricted to easy-plane magnets, we have investigated the possibility of superfluid-like spin transport in an easy-axis ferromagnet driven to a spin-torque oscillating regime. We have identified the condition for the spin torque with which the system can be stabilized to a dynamic easy-cone state that breaks the U(1) spin-rotational symmetry spontaneously. By combining the theoretical analysis and the micromagnetic simulations, we have shown that the spin current injected from one end of the ferromagnet decays algebraically, rather than exponentially, as the system length increases, whereby demonstrating that superfluid-like spin transport can be achieved in an easy-axis ferromagnet under suitable dynamic biases. We hope that our work stimulates further investigations of superfluid-like spin transport and other unconventional spin transport in various types of magnets, departing from simple easy-plane or easy-axis magnets.

\begin{acknowledgments}
We acknowledge the discussion with Yaroslav Tserkovnyak. This work was supported by Brain Pool Plus Program through the National Research Foundation of Korea funded by the Ministry of Science and ICT (NRF-2020H1D3A2A03099291), by the National Research Foundation of Korea(NRF) grant funded by the Korea government(MSIT) (NRF-2021R1C1C1006273), by the National Research Foundation of Korea funded by the Korea Government via the SRC Center for Quantum Coherence in Condensed Matter (NRF-2016R1A5A1008184), and by Basic Science Research Program through the KAIST Basic Science 4.0 Priority Research Center funded by the Ministry of Science and ICT.
\end{acknowledgments}

\bibliography{ref.bib}

\begin{thebibliography}{66}%
\makeatletter
\providecommand \@ifxundefined [1]{%
 \@ifx{#1\undefined}
}%
\providecommand \@ifnum [1]{%
 \ifnum #1\expandafter \@firstoftwo
 \else \expandafter \@secondoftwo
 \fi
}%
\providecommand \@ifx [1]{%
 \ifx #1\expandafter \@firstoftwo
 \else \expandafter \@secondoftwo
 \fi
}%
\providecommand \natexlab [1]{#1}%
\providecommand \enquote  [1]{``#1''}%
\providecommand \bibnamefont  [1]{#1}%
\providecommand \bibfnamefont [1]{#1}%
\providecommand \citenamefont [1]{#1}%
\providecommand \href@noop [0]{\@secondoftwo}%
\providecommand \href [0]{\begingroup \@sanitize@url \@href}%
\providecommand \@href[1]{\@@startlink{#1}\@@href}%
\providecommand \@@href[1]{\endgroup#1\@@endlink}%
\providecommand \@sanitize@url [0]{\catcode `\\12\catcode `\$12\catcode
  `\&12\catcode `\#12\catcode `\^12\catcode `\_12\catcode `\%12\relax}%
\providecommand \@@startlink[1]{}%
\providecommand \@@endlink[0]{}%
\providecommand \url  [0]{\begingroup\@sanitize@url \@url }%
\providecommand \@url [1]{\endgroup\@href {#1}{\urlprefix }}%
\providecommand \urlprefix  [0]{URL }%
\providecommand \Eprint [0]{\href }%
\providecommand \doibase [0]{https://doi.org/}%
\providecommand \selectlanguage [0]{\@gobble}%
\providecommand \bibinfo  [0]{\@secondoftwo}%
\providecommand \bibfield  [0]{\@secondoftwo}%
\providecommand \translation [1]{[#1]}%
\providecommand \BibitemOpen [0]{}%
\providecommand \bibitemStop [0]{}%
\providecommand \bibitemNoStop [0]{.\EOS\space}%
\providecommand \EOS [0]{\spacefactor3000\relax}%
\providecommand \BibitemShut  [1]{\csname bibitem#1\endcsname}%
\let\auto@bib@innerbib\@empty
\bibitem [{\citenamefont {\ifmmode \check{Z}\else
  \v{Z}\fi{}uti\ifmmode~\acute{c}\else \'{c}\fi{}}\ \emph
  {et~al.}(2004)\citenamefont {\ifmmode \check{Z}\else
  \v{Z}\fi{}uti\ifmmode~\acute{c}\else \'{c}\fi{}}, \citenamefont {Fabian},\
  and\ \citenamefont {Das~Sarma}}]{Igorspintronics2004}%
  \BibitemOpen
  \bibfield  {author} {\bibinfo {author} {\bibfnamefont {I.}~\bibnamefont
  {\ifmmode \check{Z}\else \v{Z}\fi{}uti\ifmmode~\acute{c}\else \'{c}\fi{}}},
  \bibinfo {author} {\bibfnamefont {J.}~\bibnamefont {Fabian}},\ and\ \bibinfo
  {author} {\bibfnamefont {S.}~\bibnamefont {Das~Sarma}},\ }\bibfield  {title}
  {\bibinfo {title} {Spintronics: Fundamentals and applications},\ }\href
  {https://doi.org/10.1103/RevModPhys.76.323} {\bibfield  {journal} {\bibinfo
  {journal} {Rev. Mod. Phys.}\ }\textbf {\bibinfo {volume} {76}},\ \bibinfo
  {pages} {323} (\bibinfo {year} {2004})}\BibitemShut {NoStop}%
\bibitem [{\citenamefont {Wolf}\ \emph {et~al.}(2001)\citenamefont {Wolf},
  \citenamefont {Awschalom}, \citenamefont {Buhrman}, \citenamefont {Daughton},
  \citenamefont {von Moln{\'a}r}, \citenamefont {Roukes}, \citenamefont
  {Chtchelkanova},\ and\ \citenamefont {Treger}}]{Wolfspintronics2001}%
  \BibitemOpen
  \bibfield  {author} {\bibinfo {author} {\bibfnamefont {S.~A.}\ \bibnamefont
  {Wolf}}, \bibinfo {author} {\bibfnamefont {D.~D.}\ \bibnamefont {Awschalom}},
  \bibinfo {author} {\bibfnamefont {R.~A.}\ \bibnamefont {Buhrman}}, \bibinfo
  {author} {\bibfnamefont {J.~M.}\ \bibnamefont {Daughton}}, \bibinfo {author}
  {\bibfnamefont {S.}~\bibnamefont {von Moln{\'a}r}}, \bibinfo {author}
  {\bibfnamefont {M.~L.}\ \bibnamefont {Roukes}}, \bibinfo {author}
  {\bibfnamefont {A.~Y.}\ \bibnamefont {Chtchelkanova}},\ and\ \bibinfo
  {author} {\bibfnamefont {D.~M.}\ \bibnamefont {Treger}},\ }\bibfield  {title}
  {\bibinfo {title} {Spintronics: A spin-based electronics vision for the
  future},\ }\href {https://doi.org/10.1126/science.1065389} {\bibfield
  {journal} {\bibinfo  {journal} {Science}\ }\textbf {\bibinfo {volume}
  {294}},\ \bibinfo {pages} {1488} (\bibinfo {year} {2001})}\BibitemShut
  {NoStop}%
\bibitem [{\citenamefont {Guo}\ \emph {et~al.}(2019)\citenamefont {Guo},
  \citenamefont {Gu}, \citenamefont {Zhu},\ and\ \citenamefont
  {Sun}}]{Guorecent2019}%
  \BibitemOpen
  \bibfield  {author} {\bibinfo {author} {\bibfnamefont {L.}~\bibnamefont
  {Guo}}, \bibinfo {author} {\bibfnamefont {X.}~\bibnamefont {Gu}}, \bibinfo
  {author} {\bibfnamefont {X.}~\bibnamefont {Zhu}},\ and\ \bibinfo {author}
  {\bibfnamefont {X.}~\bibnamefont {Sun}},\ }\bibfield  {title} {\bibinfo
  {title} {Recent advances in molecular spintronics: Multifunctional spintronic
  devices},\ }\href {https://doi.org/https://doi.org/10.1002/adma.201805355}
  {\bibfield  {journal} {\bibinfo  {journal} {Adv. Mater.}\ }\textbf {\bibinfo
  {volume} {31}},\ \bibinfo {pages} {1805355} (\bibinfo {year}
  {2019})}\BibitemShut {NoStop}%
\bibitem [{\citenamefont {Kittel}\ \emph {et~al.}(1996)\citenamefont {Kittel},
  \citenamefont {McEuen},\ and\ \citenamefont {McEuen}}]{Kittel}%
  \BibitemOpen
  \bibfield  {author} {\bibinfo {author} {\bibfnamefont {C.}~\bibnamefont
  {Kittel}}, \bibinfo {author} {\bibfnamefont {P.}~\bibnamefont {McEuen}},\
  and\ \bibinfo {author} {\bibfnamefont {P.}~\bibnamefont {McEuen}},\
  }\href@noop {} {\emph {\bibinfo {title} {Introduction to solid state
  physics}}},\ Vol.~\bibinfo {volume} {8}\ (\bibinfo  {publisher} {Wiley New
  York},\ \bibinfo {year} {1996})\BibitemShut {NoStop}%
\bibitem [{\citenamefont {Kajiwara}\ \emph {et~al.}(2010)\citenamefont
  {Kajiwara}, \citenamefont {Harii}, \citenamefont {Takahashi}, \citenamefont
  {Ohe}, \citenamefont {Uchida}, \citenamefont {Mizuguchi}, \citenamefont
  {Umezawa}, \citenamefont {Kawai}, \citenamefont {Ando}, \citenamefont
  {Takanashi}, \citenamefont {Maekawa},\ and\ \citenamefont
  {Saitoh}}]{Kajiwaratransmission2010}%
  \BibitemOpen
  \bibfield  {author} {\bibinfo {author} {\bibfnamefont {Y.}~\bibnamefont
  {Kajiwara}}, \bibinfo {author} {\bibfnamefont {K.}~\bibnamefont {Harii}},
  \bibinfo {author} {\bibfnamefont {S.}~\bibnamefont {Takahashi}}, \bibinfo
  {author} {\bibfnamefont {J.}~\bibnamefont {Ohe}}, \bibinfo {author}
  {\bibfnamefont {K.}~\bibnamefont {Uchida}}, \bibinfo {author} {\bibfnamefont
  {M.}~\bibnamefont {Mizuguchi}}, \bibinfo {author} {\bibfnamefont
  {H.}~\bibnamefont {Umezawa}}, \bibinfo {author} {\bibfnamefont
  {H.}~\bibnamefont {Kawai}}, \bibinfo {author} {\bibfnamefont
  {K.}~\bibnamefont {Ando}}, \bibinfo {author} {\bibfnamefont {K.}~\bibnamefont
  {Takanashi}}, \bibinfo {author} {\bibfnamefont {S.}~\bibnamefont {Maekawa}},\
  and\ \bibinfo {author} {\bibfnamefont {E.}~\bibnamefont {Saitoh}},\
  }\bibfield  {title} {\bibinfo {title} {Transmission of electrical signals by
  spin-wave interconversion in a magnetic insulator},\ }\href
  {https://doi.org/10.1038/nature08876} {\bibfield  {journal} {\bibinfo
  {journal} {Nature}\ }\textbf {\bibinfo {volume} {464}},\ \bibinfo {pages}
  {262} (\bibinfo {year} {2010})}\BibitemShut {NoStop}%
\bibitem [{\citenamefont {Uchida}\ \emph {et~al.}(2010)\citenamefont {Uchida},
  \citenamefont {Xiao}, \citenamefont {Adachi}, \citenamefont {Ohe},
  \citenamefont {Takahashi}, \citenamefont {Ieda}, \citenamefont {Ota},
  \citenamefont {Kajiwara}, \citenamefont {Umezawa}, \citenamefont {Kawai},
  \citenamefont {Bauer}, \citenamefont {Maekawa},\ and\ \citenamefont
  {Saitoh}}]{Uchidaspin2010}%
  \BibitemOpen
  \bibfield  {author} {\bibinfo {author} {\bibfnamefont {K.}~\bibnamefont
  {Uchida}}, \bibinfo {author} {\bibfnamefont {J.}~\bibnamefont {Xiao}},
  \bibinfo {author} {\bibfnamefont {H.}~\bibnamefont {Adachi}}, \bibinfo
  {author} {\bibfnamefont {J.}~\bibnamefont {Ohe}}, \bibinfo {author}
  {\bibfnamefont {S.}~\bibnamefont {Takahashi}}, \bibinfo {author}
  {\bibfnamefont {J.}~\bibnamefont {Ieda}}, \bibinfo {author} {\bibfnamefont
  {T.}~\bibnamefont {Ota}}, \bibinfo {author} {\bibfnamefont {Y.}~\bibnamefont
  {Kajiwara}}, \bibinfo {author} {\bibfnamefont {H.}~\bibnamefont {Umezawa}},
  \bibinfo {author} {\bibfnamefont {H.}~\bibnamefont {Kawai}}, \bibinfo
  {author} {\bibfnamefont {G.~E.~W.}\ \bibnamefont {Bauer}}, \bibinfo {author}
  {\bibfnamefont {S.}~\bibnamefont {Maekawa}},\ and\ \bibinfo {author}
  {\bibfnamefont {E.}~\bibnamefont {Saitoh}},\ }\bibfield  {title} {\bibinfo
  {title} {Spin seebeck insulator},\ }\href {https://doi.org/10.1038/nmat2856}
  {\bibfield  {journal} {\bibinfo  {journal} {Nat. Mater.}\ }\textbf {\bibinfo
  {volume} {9}},\ \bibinfo {pages} {894} (\bibinfo {year} {2010})}\BibitemShut
  {NoStop}%
\bibitem [{\citenamefont {Chumak}\ \emph {et~al.}(2015)\citenamefont {Chumak},
  \citenamefont {Vasyuchka}, \citenamefont {Serga},\ and\ \citenamefont
  {Hillebrands}}]{Chumakmagnon2015}%
  \BibitemOpen
  \bibfield  {author} {\bibinfo {author} {\bibfnamefont {A.~V.}\ \bibnamefont
  {Chumak}}, \bibinfo {author} {\bibfnamefont {V.~I.}\ \bibnamefont
  {Vasyuchka}}, \bibinfo {author} {\bibfnamefont {A.~A.}\ \bibnamefont
  {Serga}},\ and\ \bibinfo {author} {\bibfnamefont {B.}~\bibnamefont
  {Hillebrands}},\ }\bibfield  {title} {\bibinfo {title} {Magnon spintronics},\
  }\href {https://doi.org/10.1038/nphys3347} {\bibfield  {journal} {\bibinfo
  {journal} {Nat. Phys.}\ }\textbf {\bibinfo {volume} {11}},\ \bibinfo {pages}
  {453} (\bibinfo {year} {2015})}\BibitemShut {NoStop}%
\bibitem [{\citenamefont {Navabi}\ \emph {et~al.}(2019)\citenamefont {Navabi},
  \citenamefont {Liu}, \citenamefont {Upadhyaya}, \citenamefont {Murata},
  \citenamefont {Ebrahimi}, \citenamefont {Yu}, \citenamefont {Ma},
  \citenamefont {Rao}, \citenamefont {Yazdani}, \citenamefont {Montazeri},
  \citenamefont {Pan}, \citenamefont {Krivorotov}, \citenamefont {Barsukov},
  \citenamefont {Yang}, \citenamefont {Khalili~Amiri}, \citenamefont
  {Tserkovnyak},\ and\ \citenamefont {Wang}}]{Navabicontrol2019}%
  \BibitemOpen
  \bibfield  {author} {\bibinfo {author} {\bibfnamefont {A.}~\bibnamefont
  {Navabi}}, \bibinfo {author} {\bibfnamefont {Y.}~\bibnamefont {Liu}},
  \bibinfo {author} {\bibfnamefont {P.}~\bibnamefont {Upadhyaya}}, \bibinfo
  {author} {\bibfnamefont {K.}~\bibnamefont {Murata}}, \bibinfo {author}
  {\bibfnamefont {F.}~\bibnamefont {Ebrahimi}}, \bibinfo {author}
  {\bibfnamefont {G.}~\bibnamefont {Yu}}, \bibinfo {author} {\bibfnamefont
  {B.}~\bibnamefont {Ma}}, \bibinfo {author} {\bibfnamefont {Y.}~\bibnamefont
  {Rao}}, \bibinfo {author} {\bibfnamefont {M.}~\bibnamefont {Yazdani}},
  \bibinfo {author} {\bibfnamefont {M.}~\bibnamefont {Montazeri}}, \bibinfo
  {author} {\bibfnamefont {L.}~\bibnamefont {Pan}}, \bibinfo {author}
  {\bibfnamefont {I.~N.}\ \bibnamefont {Krivorotov}}, \bibinfo {author}
  {\bibfnamefont {I.}~\bibnamefont {Barsukov}}, \bibinfo {author}
  {\bibfnamefont {Q.}~\bibnamefont {Yang}}, \bibinfo {author} {\bibfnamefont
  {P.}~\bibnamefont {Khalili~Amiri}}, \bibinfo {author} {\bibfnamefont
  {Y.}~\bibnamefont {Tserkovnyak}},\ and\ \bibinfo {author} {\bibfnamefont
  {K.~L.}\ \bibnamefont {Wang}},\ }\bibfield  {title} {\bibinfo {title}
  {Control of spin-wave damping in yig using spin currents from topological
  insulators},\ }\href {https://doi.org/10.1103/PhysRevApplied.11.034046}
  {\bibfield  {journal} {\bibinfo  {journal} {Phys. Rev. Applied}\ }\textbf
  {\bibinfo {volume} {11}},\ \bibinfo {pages} {034046} (\bibinfo {year}
  {2019})}\BibitemShut {NoStop}%
\bibitem [{\citenamefont {Barsukov}\ \emph {et~al.}(2019)\citenamefont
  {Barsukov}, \citenamefont {Lee}, \citenamefont {Jara}, \citenamefont {Chen},
  \citenamefont {Gon{\c c}alves}, \citenamefont {Sha}, \citenamefont {Katine},
  \citenamefont {Arias}, \citenamefont {Ivanov},\ and\ \citenamefont
  {Krivorotov}}]{Barsukovgiant2019}%
  \BibitemOpen
  \bibfield  {author} {\bibinfo {author} {\bibfnamefont {I.}~\bibnamefont
  {Barsukov}}, \bibinfo {author} {\bibfnamefont {H.~K.}\ \bibnamefont {Lee}},
  \bibinfo {author} {\bibfnamefont {A.~A.}\ \bibnamefont {Jara}}, \bibinfo
  {author} {\bibfnamefont {Y.-J.}\ \bibnamefont {Chen}}, \bibinfo {author}
  {\bibfnamefont {A.~M.}\ \bibnamefont {Gon{\c c}alves}}, \bibinfo {author}
  {\bibfnamefont {C.}~\bibnamefont {Sha}}, \bibinfo {author} {\bibfnamefont
  {J.~A.}\ \bibnamefont {Katine}}, \bibinfo {author} {\bibfnamefont {R.~E.}\
  \bibnamefont {Arias}}, \bibinfo {author} {\bibfnamefont {B.~A.}\ \bibnamefont
  {Ivanov}},\ and\ \bibinfo {author} {\bibfnamefont {I.~N.}\ \bibnamefont
  {Krivorotov}},\ }\bibfield  {title} {\bibinfo {title} {Giant nonlinear
  damping in nanoscale ferromagnets},\ }\href
  {https://doi.org/10.1126/sciadv.aav6943} {\bibfield  {journal} {\bibinfo
  {journal} {Sci. Adv.}\ }\textbf {\bibinfo {volume} {5}},\ \bibinfo {pages}
  {eaav6943} (\bibinfo {year} {2019})}\BibitemShut {NoStop}%
\bibitem [{\citenamefont {Etesamirad}\ \emph {et~al.}(2021)\citenamefont
  {Etesamirad}, \citenamefont {Rodriguez}, \citenamefont {Bocanegra},
  \citenamefont {Verba}, \citenamefont {Katine}, \citenamefont {Krivorotov},
  \citenamefont {Tyberkevych}, \citenamefont {Ivanov},\ and\ \citenamefont
  {Barsukov}}]{Etesamiradcontrol2021}%
  \BibitemOpen
  \bibfield  {author} {\bibinfo {author} {\bibfnamefont {A.}~\bibnamefont
  {Etesamirad}}, \bibinfo {author} {\bibfnamefont {R.}~\bibnamefont
  {Rodriguez}}, \bibinfo {author} {\bibfnamefont {J.}~\bibnamefont
  {Bocanegra}}, \bibinfo {author} {\bibfnamefont {R.}~\bibnamefont {Verba}},
  \bibinfo {author} {\bibfnamefont {J.}~\bibnamefont {Katine}}, \bibinfo
  {author} {\bibfnamefont {I.~N.}\ \bibnamefont {Krivorotov}}, \bibinfo
  {author} {\bibfnamefont {V.}~\bibnamefont {Tyberkevych}}, \bibinfo {author}
  {\bibfnamefont {B.}~\bibnamefont {Ivanov}},\ and\ \bibinfo {author}
  {\bibfnamefont {I.}~\bibnamefont {Barsukov}},\ }\bibfield  {title} {\bibinfo
  {title} {Controlling magnon interaction by a nanoscale switch},\ }\href
  {https://doi.org/10.1021/acsami.1c01562} {\bibfield  {journal} {\bibinfo
  {journal} {ACS Appl. Mater. Inter.}\ }\textbf {\bibinfo {volume} {13}},\
  \bibinfo {pages} {20288} (\bibinfo {year} {2021})}\BibitemShut {NoStop}%
\bibitem [{\citenamefont {Cornelissen}\ \emph {et~al.}(2015)\citenamefont
  {Cornelissen}, \citenamefont {Liu}, \citenamefont {Duine}, \citenamefont
  {Youssef},\ and\ \citenamefont {van Wees}}]{Cornelissenlong2015}%
  \BibitemOpen
  \bibfield  {author} {\bibinfo {author} {\bibfnamefont {L.~J.}\ \bibnamefont
  {Cornelissen}}, \bibinfo {author} {\bibfnamefont {J.}~\bibnamefont {Liu}},
  \bibinfo {author} {\bibfnamefont {R.~A.}\ \bibnamefont {Duine}}, \bibinfo
  {author} {\bibfnamefont {J.~B.}\ \bibnamefont {Youssef}},\ and\ \bibinfo
  {author} {\bibfnamefont {B.~J.}\ \bibnamefont {van Wees}},\ }\bibfield
  {title} {\bibinfo {title} {Long-distance transport of magnon spin information
  in a magnetic insulator at room temperature},\ }\href
  {https://doi.org/10.1038/nphys3465} {\bibfield  {journal} {\bibinfo
  {journal} {Nat. Phys.}\ }\textbf {\bibinfo {volume} {11}},\ \bibinfo {pages}
  {1022} (\bibinfo {year} {2015})}\BibitemShut {NoStop}%
\bibitem [{\citenamefont {Cornelissen}\ \emph {et~al.}(2016)\citenamefont
  {Cornelissen}, \citenamefont {Peters}, \citenamefont {Bauer}, \citenamefont
  {Duine},\ and\ \citenamefont {van Wees}}]{Cornelissenmagnon2016}%
  \BibitemOpen
  \bibfield  {author} {\bibinfo {author} {\bibfnamefont {L.~J.}\ \bibnamefont
  {Cornelissen}}, \bibinfo {author} {\bibfnamefont {K.~J.~H.}\ \bibnamefont
  {Peters}}, \bibinfo {author} {\bibfnamefont {G.~E.~W.}\ \bibnamefont
  {Bauer}}, \bibinfo {author} {\bibfnamefont {R.~A.}\ \bibnamefont {Duine}},\
  and\ \bibinfo {author} {\bibfnamefont {B.~J.}\ \bibnamefont {van Wees}},\
  }\bibfield  {title} {\bibinfo {title} {Magnon spin transport driven by the
  magnon chemical potential in a magnetic insulator},\ }\href
  {https://doi.org/10.1103/PhysRevB.94.014412} {\bibfield  {journal} {\bibinfo
  {journal} {Phys. Rev. B}\ }\textbf {\bibinfo {volume} {94}},\ \bibinfo
  {pages} {014412} (\bibinfo {year} {2016})}\BibitemShut {NoStop}%
\bibitem [{\citenamefont {Wesenberg}\ \emph {et~al.}(2017)\citenamefont
  {Wesenberg}, \citenamefont {Liu}, \citenamefont {Balzar}, \citenamefont
  {Wu},\ and\ \citenamefont {Zink}}]{Wesenberglong2017}%
  \BibitemOpen
  \bibfield  {author} {\bibinfo {author} {\bibfnamefont {D.}~\bibnamefont
  {Wesenberg}}, \bibinfo {author} {\bibfnamefont {T.}~\bibnamefont {Liu}},
  \bibinfo {author} {\bibfnamefont {D.}~\bibnamefont {Balzar}}, \bibinfo
  {author} {\bibfnamefont {M.}~\bibnamefont {Wu}},\ and\ \bibinfo {author}
  {\bibfnamefont {B.~L.}\ \bibnamefont {Zink}},\ }\bibfield  {title} {\bibinfo
  {title} {Long-distance spin transport in a disordered magnetic insulator},\
  }\href {https://doi.org/10.1038/nphys4175} {\bibfield  {journal} {\bibinfo
  {journal} {Nat. Phys.}\ }\textbf {\bibinfo {volume} {13}},\ \bibinfo {pages}
  {987} (\bibinfo {year} {2017})}\BibitemShut {NoStop}%
\bibitem [{\citenamefont {Sonin}(2010)}]{Soninspin2010}%
  \BibitemOpen
  \bibfield  {author} {\bibinfo {author} {\bibfnamefont {E.}~\bibnamefont
  {Sonin}},\ }\bibfield  {title} {\bibinfo {title} {Spin currents and spin
  superfluidity},\ }\href {https://doi.org/10.1080/00018731003739943}
  {\bibfield  {journal} {\bibinfo  {journal} {Adv. Phys.}\ }\textbf {\bibinfo
  {volume} {59}},\ \bibinfo {pages} {181} (\bibinfo {year} {2010})}\BibitemShut
  {NoStop}%
\bibitem [{\citenamefont {Annett}\ \emph {et~al.}(2004)\citenamefont {Annett}
  \emph {et~al.}}]{Oxfordsuper2004}%
  \BibitemOpen
  \bibfield  {author} {\bibinfo {author} {\bibfnamefont {J.~F.}\ \bibnamefont
  {Annett}} \emph {et~al.},\ }\href@noop {} {\emph {\bibinfo {title}
  {Superconductivity, superfluids and condensates}}},\ Vol.~\bibinfo {volume}
  {5}\ (\bibinfo  {publisher} {Oxford University Press},\ \bibinfo {year}
  {2004})\BibitemShut {NoStop}%
\bibitem [{\citenamefont {Sonin}(2013)}]{Soninspin2013}%
  \BibitemOpen
  \bibfield  {author} {\bibinfo {author} {\bibfnamefont {E.~B.}\ \bibnamefont
  {Sonin}},\ }\bibfield  {title} {\bibinfo {title} {Spin superfluidity,
  coherent spin precession, and magnon bec},\ }\href
  {https://doi.org/10.1007/s10909-012-0802-5} {\bibfield  {journal} {\bibinfo
  {journal} {J. Low Temp. Phys.}\ }\textbf {\bibinfo {volume} {171}},\ \bibinfo
  {pages} {757} (\bibinfo {year} {2013})}\BibitemShut {NoStop}%
\bibitem [{\citenamefont {K\"onig}\ \emph {et~al.}(2001)\citenamefont
  {K\"onig}, \citenamefont {B\o{}nsager},\ and\ \citenamefont
  {MacDonald}}]{Konigdissipationless2001}%
  \BibitemOpen
  \bibfield  {author} {\bibinfo {author} {\bibfnamefont {J.}~\bibnamefont
  {K\"onig}}, \bibinfo {author} {\bibfnamefont {M.~C.}\ \bibnamefont
  {B\o{}nsager}},\ and\ \bibinfo {author} {\bibfnamefont {A.~H.}\ \bibnamefont
  {MacDonald}},\ }\bibfield  {title} {\bibinfo {title} {Dissipationless spin
  transport in thin film ferromagnets},\ }\href
  {https://doi.org/10.1103/PhysRevLett.87.187202} {\bibfield  {journal}
  {\bibinfo  {journal} {Phys. Rev. Lett.}\ }\textbf {\bibinfo {volume} {87}},\
  \bibinfo {pages} {187202} (\bibinfo {year} {2001})}\BibitemShut {NoStop}%
\bibitem [{\citenamefont {Chen}\ \emph {et~al.}(2014)\citenamefont {Chen},
  \citenamefont {Kent}, \citenamefont {MacDonald},\ and\ \citenamefont
  {Sodemann}}]{Chennonlocal2014}%
  \BibitemOpen
  \bibfield  {author} {\bibinfo {author} {\bibfnamefont {H.}~\bibnamefont
  {Chen}}, \bibinfo {author} {\bibfnamefont {A.~D.}\ \bibnamefont {Kent}},
  \bibinfo {author} {\bibfnamefont {A.~H.}\ \bibnamefont {MacDonald}},\ and\
  \bibinfo {author} {\bibfnamefont {I.}~\bibnamefont {Sodemann}},\ }\bibfield
  {title} {\bibinfo {title} {Nonlocal transport mediated by spin
  supercurrents},\ }\href {https://doi.org/10.1103/PhysRevB.90.220401}
  {\bibfield  {journal} {\bibinfo  {journal} {Phys. Rev. B}\ }\textbf {\bibinfo
  {volume} {90}},\ \bibinfo {pages} {220401} (\bibinfo {year}
  {2014})}\BibitemShut {NoStop}%
\bibitem [{\citenamefont {Chen}\ and\ \citenamefont
  {Sigrist}(2014)}]{Chenspin2014}%
  \BibitemOpen
  \bibfield  {author} {\bibinfo {author} {\bibfnamefont {W.}~\bibnamefont
  {Chen}}\ and\ \bibinfo {author} {\bibfnamefont {M.}~\bibnamefont {Sigrist}},\
  }\bibfield  {title} {\bibinfo {title} {Spin superfluidity in coplanar
  multiferroics},\ }\href {https://doi.org/10.1103/PhysRevB.89.024511}
  {\bibfield  {journal} {\bibinfo  {journal} {Phys. Rev. B}\ }\textbf {\bibinfo
  {volume} {89}},\ \bibinfo {pages} {024511} (\bibinfo {year}
  {2014})}\BibitemShut {NoStop}%
\bibitem [{\citenamefont {Skarsv\aa{}g}\ \emph {et~al.}(2015)\citenamefont
  {Skarsv\aa{}g}, \citenamefont {Holmqvist},\ and\ \citenamefont
  {Brataas}}]{Hansspin2015}%
  \BibitemOpen
  \bibfield  {author} {\bibinfo {author} {\bibfnamefont {H.}~\bibnamefont
  {Skarsv\aa{}g}}, \bibinfo {author} {\bibfnamefont {C.}~\bibnamefont
  {Holmqvist}},\ and\ \bibinfo {author} {\bibfnamefont {A.}~\bibnamefont
  {Brataas}},\ }\bibfield  {title} {\bibinfo {title} {Spin superfluidity and
  long-range transport in thin-film ferromagnets},\ }\href
  {https://doi.org/10.1103/PhysRevLett.115.237201} {\bibfield  {journal}
  {\bibinfo  {journal} {Phys. Rev. Lett.}\ }\textbf {\bibinfo {volume} {115}},\
  \bibinfo {pages} {237201} (\bibinfo {year} {2015})}\BibitemShut {NoStop}%
\bibitem [{\citenamefont {Chen}\ and\ \citenamefont
  {Sigrist}(2015)}]{Chendissi2015}%
  \BibitemOpen
  \bibfield  {author} {\bibinfo {author} {\bibfnamefont {W.}~\bibnamefont
  {Chen}}\ and\ \bibinfo {author} {\bibfnamefont {M.}~\bibnamefont {Sigrist}},\
  }\bibfield  {title} {\bibinfo {title} {Dissipationless multiferroic
  magnonics},\ }\href {https://doi.org/10.1103/PhysRevLett.114.157203}
  {\bibfield  {journal} {\bibinfo  {journal} {Phys. Rev. Lett.}\ }\textbf
  {\bibinfo {volume} {114}},\ \bibinfo {pages} {157203} (\bibinfo {year}
  {2015})}\BibitemShut {NoStop}%
\bibitem [{\citenamefont {Linder}\ and\ \citenamefont
  {Robinson}(2015)}]{Lindersuper2015}%
  \BibitemOpen
  \bibfield  {author} {\bibinfo {author} {\bibfnamefont {J.}~\bibnamefont
  {Linder}}\ and\ \bibinfo {author} {\bibfnamefont {J.~W.~A.}\ \bibnamefont
  {Robinson}},\ }\bibfield  {title} {\bibinfo {title} {Superconducting
  spintronics},\ }\href {https://doi.org/10.1038/nphys3242} {\bibfield
  {journal} {\bibinfo  {journal} {Nat. Phys.}\ }\textbf {\bibinfo {volume}
  {11}},\ \bibinfo {pages} {307} (\bibinfo {year} {2015})}\BibitemShut
  {NoStop}%
\bibitem [{\citenamefont {Flebus}\ \emph {et~al.}(2016)\citenamefont {Flebus},
  \citenamefont {Bender}, \citenamefont {Tserkovnyak},\ and\ \citenamefont
  {Duine}}]{Flebustwofluid2016}%
  \BibitemOpen
  \bibfield  {author} {\bibinfo {author} {\bibfnamefont {B.}~\bibnamefont
  {Flebus}}, \bibinfo {author} {\bibfnamefont {S.~A.}\ \bibnamefont {Bender}},
  \bibinfo {author} {\bibfnamefont {Y.}~\bibnamefont {Tserkovnyak}},\ and\
  \bibinfo {author} {\bibfnamefont {R.~A.}\ \bibnamefont {Duine}},\ }\bibfield
  {title} {\bibinfo {title} {Two-fluid theory for spin superfluidity in
  magnetic insulators},\ }\href
  {https://doi.org/10.1103/PhysRevLett.116.117201} {\bibfield  {journal}
  {\bibinfo  {journal} {Phys. Rev. Lett.}\ }\textbf {\bibinfo {volume} {116}},\
  \bibinfo {pages} {117201} (\bibinfo {year} {2016})}\BibitemShut {NoStop}%
\bibitem [{\citenamefont {Kim}\ and\ \citenamefont
  {Tserkovnyak}(2016)}]{Kiminteraction2016}%
  \BibitemOpen
  \bibfield  {author} {\bibinfo {author} {\bibfnamefont {S.~K.}\ \bibnamefont
  {Kim}}\ and\ \bibinfo {author} {\bibfnamefont {Y.}~\bibnamefont
  {Tserkovnyak}},\ }\bibfield  {title} {\bibinfo {title} {Interaction between a
  domain wall and spin supercurrent in easy-cone magnets},\ }\href
  {https://doi.org/10.1103/PhysRevB.94.220404} {\bibfield  {journal} {\bibinfo
  {journal} {Phys. Rev. B}\ }\textbf {\bibinfo {volume} {94}},\ \bibinfo
  {pages} {220404} (\bibinfo {year} {2016})}\BibitemShut {NoStop}%
\bibitem [{\citenamefont {Cheng}\ \emph {et~al.}(2016)\citenamefont {Cheng},
  \citenamefont {Xiao},\ and\ \citenamefont {Brataas}}]{Chengtera2016}%
  \BibitemOpen
  \bibfield  {author} {\bibinfo {author} {\bibfnamefont {R.}~\bibnamefont
  {Cheng}}, \bibinfo {author} {\bibfnamefont {D.}~\bibnamefont {Xiao}},\ and\
  \bibinfo {author} {\bibfnamefont {A.}~\bibnamefont {Brataas}},\ }\bibfield
  {title} {\bibinfo {title} {Terahertz antiferromagnetic spin hall
  nano-oscillator},\ }\href {https://doi.org/10.1103/PhysRevLett.116.207603}
  {\bibfield  {journal} {\bibinfo  {journal} {Phys. Rev. Lett.}\ }\textbf
  {\bibinfo {volume} {116}},\ \bibinfo {pages} {207603} (\bibinfo {year}
  {2016})}\BibitemShut {NoStop}%
\bibitem [{\citenamefont {Qaiumzadeh}\ \emph {et~al.}(2017)\citenamefont
  {Qaiumzadeh}, \citenamefont {Skarsv\aa{}g}, \citenamefont {Holmqvist},\ and\
  \citenamefont {Brataas}}]{Qaiumzadehspin2017}%
  \BibitemOpen
  \bibfield  {author} {\bibinfo {author} {\bibfnamefont {A.}~\bibnamefont
  {Qaiumzadeh}}, \bibinfo {author} {\bibfnamefont {H.}~\bibnamefont
  {Skarsv\aa{}g}}, \bibinfo {author} {\bibfnamefont {C.}~\bibnamefont
  {Holmqvist}},\ and\ \bibinfo {author} {\bibfnamefont {A.}~\bibnamefont
  {Brataas}},\ }\bibfield  {title} {\bibinfo {title} {Spin superfluidity in
  biaxial antiferromagnetic insulators},\ }\href
  {https://doi.org/10.1103/PhysRevLett.118.137201} {\bibfield  {journal}
  {\bibinfo  {journal} {Phys. Rev. Lett.}\ }\textbf {\bibinfo {volume} {118}},\
  \bibinfo {pages} {137201} (\bibinfo {year} {2017})}\BibitemShut {NoStop}%
\bibitem [{\citenamefont {Takei}\ \emph {et~al.}(2014)\citenamefont {Takei},
  \citenamefont {Halperin}, \citenamefont {Yacoby},\ and\ \citenamefont
  {Tserkovnyak}}]{Takeiafm2014}%
  \BibitemOpen
  \bibfield  {author} {\bibinfo {author} {\bibfnamefont {S.}~\bibnamefont
  {Takei}}, \bibinfo {author} {\bibfnamefont {B.~I.}\ \bibnamefont {Halperin}},
  \bibinfo {author} {\bibfnamefont {A.}~\bibnamefont {Yacoby}},\ and\ \bibinfo
  {author} {\bibfnamefont {Y.}~\bibnamefont {Tserkovnyak}},\ }\bibfield
  {title} {\bibinfo {title} {Superfluid spin transport through
  antiferromagnetic insulators},\ }\href
  {https://doi.org/10.1103/PhysRevB.90.094408} {\bibfield  {journal} {\bibinfo
  {journal} {Phys. Rev. B}\ }\textbf {\bibinfo {volume} {90}},\ \bibinfo
  {pages} {094408} (\bibinfo {year} {2014})}\BibitemShut {NoStop}%
\bibitem [{\citenamefont {Takei}\ and\ \citenamefont
  {Tserkovnyak}(2014)}]{Takeifm2014}%
  \BibitemOpen
  \bibfield  {author} {\bibinfo {author} {\bibfnamefont {S.}~\bibnamefont
  {Takei}}\ and\ \bibinfo {author} {\bibfnamefont {Y.}~\bibnamefont
  {Tserkovnyak}},\ }\bibfield  {title} {\bibinfo {title} {Superfluid spin
  transport through easy-plane ferromagnetic insulators},\ }\href
  {https://doi.org/10.1103/PhysRevLett.112.227201} {\bibfield  {journal}
  {\bibinfo  {journal} {Phys. Rev. Lett.}\ }\textbf {\bibinfo {volume} {112}},\
  \bibinfo {pages} {227201} (\bibinfo {year} {2014})}\BibitemShut {NoStop}%
\bibitem [{\citenamefont {Takei}\ and\ \citenamefont
  {Tserkovnyak}(2015)}]{Takeinonlocal2015}%
  \BibitemOpen
  \bibfield  {author} {\bibinfo {author} {\bibfnamefont {S.}~\bibnamefont
  {Takei}}\ and\ \bibinfo {author} {\bibfnamefont {Y.}~\bibnamefont
  {Tserkovnyak}},\ }\bibfield  {title} {\bibinfo {title} {Nonlocal
  magnetoresistance mediated by spin superfluidity},\ }\href
  {https://doi.org/10.1103/PhysRevLett.115.156604} {\bibfield  {journal}
  {\bibinfo  {journal} {Phys. Rev. Lett.}\ }\textbf {\bibinfo {volume} {115}},\
  \bibinfo {pages} {156604} (\bibinfo {year} {2015})}\BibitemShut {NoStop}%
\bibitem [{\citenamefont {Sonin}(2019)}]{Soninsuperfluid2019}%
  \BibitemOpen
  \bibfield  {author} {\bibinfo {author} {\bibfnamefont {E.~B.}\ \bibnamefont
  {Sonin}},\ }\bibfield  {title} {\bibinfo {title} {Superfluid spin transport
  in ferro- and antiferromagnets},\ }\href
  {https://doi.org/10.1103/PhysRevB.99.104423} {\bibfield  {journal} {\bibinfo
  {journal} {Phys. Rev. B}\ }\textbf {\bibinfo {volume} {99}},\ \bibinfo
  {pages} {104423} (\bibinfo {year} {2019})}\BibitemShut {NoStop}%
\bibitem [{\citenamefont {Slonczewski}(1996)}]{slonczewskicurrent1996}%
  \BibitemOpen
  \bibfield  {author} {\bibinfo {author} {\bibfnamefont {J.}~\bibnamefont
  {Slonczewski}},\ }\bibfield  {title} {\bibinfo {title} {Current-driven
  excitation of magnetic multilayers},\ }\href
  {https://doi.org/https://doi.org/10.1016/0304-8853(96)00062-5} {\bibfield
  {journal} {\bibinfo  {journal} {J. Magn. Magn. Mater}\ }\textbf {\bibinfo
  {volume} {159}},\ \bibinfo {pages} {L1} (\bibinfo {year} {1996})}\BibitemShut
  {NoStop}%
\bibitem [{\citenamefont {Ralph}\ and\ \citenamefont
  {Stiles}(2008{\natexlab{a}})}]{Ralphspin2008}%
  \BibitemOpen
  \bibfield  {author} {\bibinfo {author} {\bibfnamefont {D.}~\bibnamefont
  {Ralph}}\ and\ \bibinfo {author} {\bibfnamefont {M.}~\bibnamefont {Stiles}},\
  }\bibfield  {title} {\bibinfo {title} {Spin transfer torques},\ }\href
  {https://doi.org/https://doi.org/10.1016/j.jmmm.2007.12.019} {\bibfield
  {journal} {\bibinfo  {journal} {J. Magn. Magn. Mater}\ }\textbf {\bibinfo
  {volume} {320}},\ \bibinfo {pages} {1190} (\bibinfo {year}
  {2008}{\natexlab{a}})}\BibitemShut {NoStop}%
\bibitem [{\citenamefont {Tserkovnyak}\ \emph {et~al.}(2008)\citenamefont
  {Tserkovnyak}, \citenamefont {Brataas},\ and\ \citenamefont
  {Bauer}}]{Yaroslavtheory2008}%
  \BibitemOpen
  \bibfield  {author} {\bibinfo {author} {\bibfnamefont {Y.}~\bibnamefont
  {Tserkovnyak}}, \bibinfo {author} {\bibfnamefont {A.}~\bibnamefont
  {Brataas}},\ and\ \bibinfo {author} {\bibfnamefont {G.~E.}\ \bibnamefont
  {Bauer}},\ }\bibfield  {title} {\bibinfo {title} {Theory of current-driven
  magnetization dynamics in inhomogeneous ferromagnets},\ }\href
  {https://doi.org/https://doi.org/10.1016/j.jmmm.2007.12.012} {\bibfield
  {journal} {\bibinfo  {journal} {J. Magn. Magn. Mater}\ }\textbf {\bibinfo
  {volume} {320}},\ \bibinfo {pages} {1282} (\bibinfo {year}
  {2008})}\BibitemShut {NoStop}%
\bibitem [{\citenamefont {Jia}\ \emph {et~al.}(2011)\citenamefont {Jia},
  \citenamefont {Liu}, \citenamefont {Xia},\ and\ \citenamefont
  {Bauer}}]{Xingtaospin2011}%
  \BibitemOpen
  \bibfield  {author} {\bibinfo {author} {\bibfnamefont {X.}~\bibnamefont
  {Jia}}, \bibinfo {author} {\bibfnamefont {K.}~\bibnamefont {Liu}}, \bibinfo
  {author} {\bibfnamefont {K.}~\bibnamefont {Xia}},\ and\ \bibinfo {author}
  {\bibfnamefont {G.~E.~W.}\ \bibnamefont {Bauer}},\ }\bibfield  {title}
  {\bibinfo {title} {Spin transfer torque on magnetic insulators},\ }\href
  {https://doi.org/10.1209/0295-5075/96/17005} {\bibfield  {journal} {\bibinfo
  {journal} {Eur. Phys. Lett}\ }\textbf {\bibinfo {volume} {96}},\ \bibinfo
  {pages} {17005} (\bibinfo {year} {2011})}\BibitemShut {NoStop}%
\bibitem [{\citenamefont {Duan}\ \emph {et~al.}(2014)\citenamefont {Duan},
  \citenamefont {Smith}, \citenamefont {Yang}, \citenamefont {Youngblood},
  \citenamefont {Lindner}, \citenamefont {Demidov}, \citenamefont
  {Demokritov},\ and\ \citenamefont {Krivorotov}}]{Duannanowire2014}%
  \BibitemOpen
  \bibfield  {author} {\bibinfo {author} {\bibfnamefont {Z.}~\bibnamefont
  {Duan}}, \bibinfo {author} {\bibfnamefont {A.}~\bibnamefont {Smith}},
  \bibinfo {author} {\bibfnamefont {L.}~\bibnamefont {Yang}}, \bibinfo {author}
  {\bibfnamefont {B.}~\bibnamefont {Youngblood}}, \bibinfo {author}
  {\bibfnamefont {J.}~\bibnamefont {Lindner}}, \bibinfo {author} {\bibfnamefont
  {V.~E.}\ \bibnamefont {Demidov}}, \bibinfo {author} {\bibfnamefont {S.~O.}\
  \bibnamefont {Demokritov}},\ and\ \bibinfo {author} {\bibfnamefont {I.~N.}\
  \bibnamefont {Krivorotov}},\ }\bibfield  {title} {\bibinfo {title} {Nanowire
  spin torque oscillator driven by spin orbit torques},\ }\href
  {https://doi.org/10.1038/ncomms6616} {\bibfield  {journal} {\bibinfo
  {journal} {Nat. Commun.}\ }\textbf {\bibinfo {volume} {5}},\ \bibinfo {pages}
  {5616} (\bibinfo {year} {2014})}\BibitemShut {NoStop}%
\bibitem [{\citenamefont {Tserkovnyak}(2018)}]{Yaroslavpers2018}%
  \BibitemOpen
  \bibfield  {author} {\bibinfo {author} {\bibfnamefont {Y.}~\bibnamefont
  {Tserkovnyak}},\ }\bibfield  {title} {\bibinfo {title} {Perspective: (beyond)
  spin transport in insulators},\ }\href {https://doi.org/10.1063/1.5054123}
  {\bibfield  {journal} {\bibinfo  {journal} {J. Appl. Phys}\ }\textbf
  {\bibinfo {volume} {124}},\ \bibinfo {pages} {190901} (\bibinfo {year}
  {2018})}\BibitemShut {NoStop}%
\bibitem [{\citenamefont {Flebus}\ \emph {et~al.}(2020)\citenamefont {Flebus},
  \citenamefont {Duine},\ and\ \citenamefont {Hurst}}]{Flebusnon2020}%
  \BibitemOpen
  \bibfield  {author} {\bibinfo {author} {\bibfnamefont {B.}~\bibnamefont
  {Flebus}}, \bibinfo {author} {\bibfnamefont {R.~A.}\ \bibnamefont {Duine}},\
  and\ \bibinfo {author} {\bibfnamefont {H.~M.}\ \bibnamefont {Hurst}},\
  }\bibfield  {title} {\bibinfo {title} {Non-hermitian topology of
  one-dimensional spin-torque oscillator arrays},\ }\href
  {https://doi.org/10.1103/PhysRevB.102.180408} {\bibfield  {journal} {\bibinfo
   {journal} {Phys. Rev. B}\ }\textbf {\bibinfo {volume} {102}},\ \bibinfo
  {pages} {180408} (\bibinfo {year} {2020})}\BibitemShut {NoStop}%
\bibitem [{\citenamefont {Kato}\ \emph {et~al.}(2004)\citenamefont {Kato},
  \citenamefont {Myers}, \citenamefont {Gossard},\ and\ \citenamefont
  {Awschalom}}]{Katoobser2004}%
  \BibitemOpen
  \bibfield  {author} {\bibinfo {author} {\bibfnamefont {Y.~K.}\ \bibnamefont
  {Kato}}, \bibinfo {author} {\bibfnamefont {R.~C.}\ \bibnamefont {Myers}},
  \bibinfo {author} {\bibfnamefont {A.~C.}\ \bibnamefont {Gossard}},\ and\
  \bibinfo {author} {\bibfnamefont {D.~D.}\ \bibnamefont {Awschalom}},\
  }\bibfield  {title} {\bibinfo {title} {Observation of the spin hall effect in
  semiconductors},\ }\href {https://doi.org/10.1126/science.1105514} {\bibfield
   {journal} {\bibinfo  {journal} {Science}\ }\textbf {\bibinfo {volume}
  {306}},\ \bibinfo {pages} {1910} (\bibinfo {year} {2004})}\BibitemShut
  {NoStop}%
\bibitem [{\citenamefont {Tserkovnyak}\ and\ \citenamefont
  {Bender}(2014)}]{Yaroslavspin2014}%
  \BibitemOpen
  \bibfield  {author} {\bibinfo {author} {\bibfnamefont {Y.}~\bibnamefont
  {Tserkovnyak}}\ and\ \bibinfo {author} {\bibfnamefont {S.~A.}\ \bibnamefont
  {Bender}},\ }\bibfield  {title} {\bibinfo {title} {Spin hall phenomenology of
  magnetic dynamics},\ }\href {https://doi.org/10.1103/PhysRevB.90.014428}
  {\bibfield  {journal} {\bibinfo  {journal} {Phys. Rev. B}\ }\textbf {\bibinfo
  {volume} {90}},\ \bibinfo {pages} {014428} (\bibinfo {year}
  {2014})}\BibitemShut {NoStop}%
\bibitem [{\citenamefont {Sinova}\ \emph {et~al.}(2015)\citenamefont {Sinova},
  \citenamefont {Valenzuela}, \citenamefont {Wunderlich}, \citenamefont
  {Back},\ and\ \citenamefont {Jungwirth}}]{Sinovaspin2015}%
  \BibitemOpen
  \bibfield  {author} {\bibinfo {author} {\bibfnamefont {J.}~\bibnamefont
  {Sinova}}, \bibinfo {author} {\bibfnamefont {S.~O.}\ \bibnamefont
  {Valenzuela}}, \bibinfo {author} {\bibfnamefont {J.}~\bibnamefont
  {Wunderlich}}, \bibinfo {author} {\bibfnamefont {C.~H.}\ \bibnamefont
  {Back}},\ and\ \bibinfo {author} {\bibfnamefont {T.}~\bibnamefont
  {Jungwirth}},\ }\bibfield  {title} {\bibinfo {title} {Spin hall effects},\
  }\href {https://doi.org/10.1103/RevModPhys.87.1213} {\bibfield  {journal}
  {\bibinfo  {journal} {Rev. Mod. Phys.}\ }\textbf {\bibinfo {volume} {87}},\
  \bibinfo {pages} {1213} (\bibinfo {year} {2015})}\BibitemShut {NoStop}%
\bibitem [{\citenamefont {Tserkovnyak}\ \emph {et~al.}(2002)\citenamefont
  {Tserkovnyak}, \citenamefont {Brataas},\ and\ \citenamefont
  {Bauer}}]{Yaroslavenhanced2002}%
  \BibitemOpen
  \bibfield  {author} {\bibinfo {author} {\bibfnamefont {Y.}~\bibnamefont
  {Tserkovnyak}}, \bibinfo {author} {\bibfnamefont {A.}~\bibnamefont
  {Brataas}},\ and\ \bibinfo {author} {\bibfnamefont {G.~E.~W.}\ \bibnamefont
  {Bauer}},\ }\bibfield  {title} {\bibinfo {title} {Enhanced gilbert damping in
  thin ferromagnetic films},\ }\href
  {https://doi.org/10.1103/PhysRevLett.88.117601} {\bibfield  {journal}
  {\bibinfo  {journal} {Phys. Rev. Lett.}\ }\textbf {\bibinfo {volume} {88}},\
  \bibinfo {pages} {117601} (\bibinfo {year} {2002})}\BibitemShut {NoStop}%
\bibitem [{\citenamefont {Tserkovnyak}\ \emph {et~al.}(2005)\citenamefont
  {Tserkovnyak}, \citenamefont {Brataas}, \citenamefont {Bauer},\ and\
  \citenamefont {Halperin}}]{Yaroslavnonlocal2005}%
  \BibitemOpen
  \bibfield  {author} {\bibinfo {author} {\bibfnamefont {Y.}~\bibnamefont
  {Tserkovnyak}}, \bibinfo {author} {\bibfnamefont {A.}~\bibnamefont
  {Brataas}}, \bibinfo {author} {\bibfnamefont {G.~E.~W.}\ \bibnamefont
  {Bauer}},\ and\ \bibinfo {author} {\bibfnamefont {B.~I.}\ \bibnamefont
  {Halperin}},\ }\bibfield  {title} {\bibinfo {title} {Nonlocal magnetization
  dynamics in ferromagnetic heterostructures},\ }\href
  {https://doi.org/10.1103/RevModPhys.77.1375} {\bibfield  {journal} {\bibinfo
  {journal} {Rev. Mod. Phys.}\ }\textbf {\bibinfo {volume} {77}},\ \bibinfo
  {pages} {1375} (\bibinfo {year} {2005})}\BibitemShut {NoStop}%
\bibitem [{\citenamefont {Landau}\ and\ \citenamefont
  {Lifshitz}(1976)}]{Quantum}%
  \BibitemOpen
  \bibfield  {author} {\bibinfo {author} {\bibfnamefont {L.~D.}\ \bibnamefont
  {Landau}}\ and\ \bibinfo {author} {\bibfnamefont {E.~M.}\ \bibnamefont
  {Lifshitz}},\ }\href@noop {} {\emph {\bibinfo {title} {Quantum Mechanics 3rd
  ed.}}}\ (\bibinfo {address} {Butterworth-Heinemann, Oxford},\ \bibinfo {year}
  {1976})\BibitemShut {NoStop}%
\bibitem [{\citenamefont {Gilbert}(2004)}]{GilbertIEEE2004}%
  \BibitemOpen
  \bibfield  {author} {\bibinfo {author} {\bibfnamefont {T.}~\bibnamefont
  {Gilbert}},\ }\bibfield  {title} {\bibinfo {title} {A phenomenological theory
  of damping in ferromagnetic materials},\ }\href
  {https://doi.org/10.1109/TMAG.2004.836740} {\bibfield  {journal} {\bibinfo
  {journal} {IEEE Trans. Magn.}\ }\textbf {\bibinfo {volume} {40}},\ \bibinfo
  {pages} {3443} (\bibinfo {year} {2004})}\BibitemShut {NoStop}%
\bibitem [{\citenamefont {Matsumoto}\ \emph {et~al.}(2015)\citenamefont
  {Matsumoto}, \citenamefont {Arai}, \citenamefont {Yuasa},\ and\ \citenamefont
  {Imamura}}]{Riespin2015}%
  \BibitemOpen
  \bibfield  {author} {\bibinfo {author} {\bibfnamefont {R.}~\bibnamefont
  {Matsumoto}}, \bibinfo {author} {\bibfnamefont {H.}~\bibnamefont {Arai}},
  \bibinfo {author} {\bibfnamefont {S.}~\bibnamefont {Yuasa}},\ and\ \bibinfo
  {author} {\bibfnamefont {H.}~\bibnamefont {Imamura}},\ }\bibfield  {title}
  {\bibinfo {title} {Spin-transfer-torque switching in a spin-valve nanopillar
  with a conically magnetized free layer},\ }\href
  {https://doi.org/10.7567/apex.8.063007} {\bibfield  {journal} {\bibinfo
  {journal} {Appl. Phys. Exp.}\ }\textbf {\bibinfo {volume} {8}},\ \bibinfo
  {pages} {063007} (\bibinfo {year} {2015})}\BibitemShut {NoStop}%
\bibitem [{\citenamefont {Shaw}\ \emph {et~al.}(2015)\citenamefont {Shaw},
  \citenamefont {Nembach}, \citenamefont {Weiler}, \citenamefont {Silva},
  \citenamefont {Schoen}, \citenamefont {Sun},\ and\ \citenamefont
  {Worledge}}]{Shawperp2015}%
  \BibitemOpen
  \bibfield  {author} {\bibinfo {author} {\bibfnamefont {J.~M.}\ \bibnamefont
  {Shaw}}, \bibinfo {author} {\bibfnamefont {H.~T.}\ \bibnamefont {Nembach}},
  \bibinfo {author} {\bibfnamefont {M.}~\bibnamefont {Weiler}}, \bibinfo
  {author} {\bibfnamefont {T.~J.}\ \bibnamefont {Silva}}, \bibinfo {author}
  {\bibfnamefont {M.}~\bibnamefont {Schoen}}, \bibinfo {author} {\bibfnamefont
  {J.~Z.}\ \bibnamefont {Sun}},\ and\ \bibinfo {author} {\bibfnamefont {D.~C.}\
  \bibnamefont {Worledge}},\ }\bibfield  {title} {\bibinfo {title}
  {Perpendicular magnetic anisotropy and easy cone state in
  {Ta}/{Co}60{Fe}20{B}20/{MgO}},\ }\href
  {https://doi.org/10.1109/LMAG.2015.2438773} {\bibfield  {journal} {\bibinfo
  {journal} {IEEE Magn. Lett}\ }\textbf {\bibinfo {volume} {6}},\ \bibinfo
  {pages} {1} (\bibinfo {year} {2015})}\BibitemShut {NoStop}%
\bibitem [{\citenamefont {Jang}\ \emph {et~al.}(2016)\citenamefont {Jang},
  \citenamefont {Lee},\ and\ \citenamefont {Lee}}]{Jangspin2016}%
  \BibitemOpen
  \bibfield  {author} {\bibinfo {author} {\bibfnamefont {P.-H.}\ \bibnamefont
  {Jang}}, \bibinfo {author} {\bibfnamefont {S.-W.}\ \bibnamefont {Lee}},\ and\
  \bibinfo {author} {\bibfnamefont {K.-J.}\ \bibnamefont {Lee}},\ }\bibfield
  {title} {\bibinfo {title} {Spin-transfer-torque-induced zero-field microwave
  oscillator using a magnetic easy cone state},\ }\href
  {https://doi.org/https://doi.org/10.1016/j.cap.2016.09.004} {\bibfield
  {journal} {\bibinfo  {journal} {Curr. Appl. Phys.}\ }\textbf {\bibinfo
  {volume} {16}},\ \bibinfo {pages} {1550} (\bibinfo {year}
  {2016})}\BibitemShut {NoStop}%
\bibitem [{\citenamefont {Jang}\ \emph {et~al.}(2019)\citenamefont {Jang},
  \citenamefont {Oh}, \citenamefont {Kim},\ and\ \citenamefont
  {Lee}}]{Jangdomain2019}%
  \BibitemOpen
  \bibfield  {author} {\bibinfo {author} {\bibfnamefont {P.-H.}\ \bibnamefont
  {Jang}}, \bibinfo {author} {\bibfnamefont {S.-H.}\ \bibnamefont {Oh}},
  \bibinfo {author} {\bibfnamefont {S.~K.}\ \bibnamefont {Kim}},\ and\ \bibinfo
  {author} {\bibfnamefont {K.-J.}\ \bibnamefont {Lee}},\ }\bibfield  {title}
  {\bibinfo {title} {Domain wall dynamics in easy-cone magnets},\ }\href
  {https://doi.org/10.1103/PhysRevB.99.024424} {\bibfield  {journal} {\bibinfo
  {journal} {Phys. Rev. B}\ }\textbf {\bibinfo {volume} {99}},\ \bibinfo
  {pages} {024424} (\bibinfo {year} {2019})}\BibitemShut {NoStop}%
\bibitem [{\citenamefont {Gweon}\ \emph {et~al.}(2020)\citenamefont {Gweon},
  \citenamefont {Park}, \citenamefont {Kim}, \citenamefont {Lee},\ and\
  \citenamefont {Lim}}]{Gweonintrinsic2020}%
  \BibitemOpen
  \bibfield  {author} {\bibinfo {author} {\bibfnamefont {H.~K.}\ \bibnamefont
  {Gweon}}, \bibinfo {author} {\bibfnamefont {H.-J.}\ \bibnamefont {Park}},
  \bibinfo {author} {\bibfnamefont {K.-W.}\ \bibnamefont {Kim}}, \bibinfo
  {author} {\bibfnamefont {K.-J.}\ \bibnamefont {Lee}},\ and\ \bibinfo {author}
  {\bibfnamefont {S.~H.}\ \bibnamefont {Lim}},\ }\bibfield  {title} {\bibinfo
  {title} {Intrinsic origin of interfacial second-order magnetic anisotropy in
  ferromagnet/normal metal heterostructures},\ }\href
  {https://doi.org/10.1038/s41427-020-0205-z} {\bibfield  {journal} {\bibinfo
  {journal} {NPG. Asia. Mater.}\ }\textbf {\bibinfo {volume} {12}},\ \bibinfo
  {pages} {23} (\bibinfo {year} {2020})}\BibitemShut {NoStop}%
\bibitem [{\citenamefont {Gweon}\ and\ \citenamefont
  {Lim}(2020)}]{Gweonevolution2020}%
  \BibitemOpen
  \bibfield  {author} {\bibinfo {author} {\bibfnamefont {H.~K.}\ \bibnamefont
  {Gweon}}\ and\ \bibinfo {author} {\bibfnamefont {S.~H.}\ \bibnamefont
  {Lim}},\ }\bibfield  {title} {\bibinfo {title} {Evolution of strong
  second-order magnetic anisotropy in pt/co/mgo trilayers by post-annealing},\
  }\href {https://doi.org/10.1063/5.0018924} {\bibfield  {journal} {\bibinfo
  {journal} {Appl. Phys. Lett.}\ }\textbf {\bibinfo {volume} {117}},\ \bibinfo
  {pages} {082403} (\bibinfo {year} {2020})}\BibitemShut {NoStop}%
\bibitem [{\citenamefont {Ralph}\ and\ \citenamefont
  {Stiles}(2008{\natexlab{b}})}]{Ralphstt}%
  \BibitemOpen
  \bibfield  {author} {\bibinfo {author} {\bibfnamefont {D.}~\bibnamefont
  {Ralph}}\ and\ \bibinfo {author} {\bibfnamefont {M.}~\bibnamefont {Stiles}},\
  }\bibfield  {title} {\bibinfo {title} {Spin transfer torques},\ }\href
  {https://doi.org/https://doi.org/10.1016/j.jmmm.2007.12.019} {\bibfield
  {journal} {\bibinfo  {journal} {J. Magn. Magn. Mater}\ }\textbf {\bibinfo
  {volume} {320}},\ \bibinfo {pages} {1190} (\bibinfo {year}
  {2008}{\natexlab{b}})}\BibitemShut {NoStop}%
\bibitem [{\citenamefont {Manchon}\ \emph {et~al.}(2019)\citenamefont
  {Manchon}, \citenamefont {\ifmmode~\check{Z}\else \v{Z}\fi{}elezn\'y},
  \citenamefont {Miron}, \citenamefont {Jungwirth}, \citenamefont {Sinova},
  \citenamefont {Thiaville}, \citenamefont {Garello},\ and\ \citenamefont
  {Gambardella}}]{Manchonsot}%
  \BibitemOpen
  \bibfield  {author} {\bibinfo {author} {\bibfnamefont {A.}~\bibnamefont
  {Manchon}}, \bibinfo {author} {\bibfnamefont {J.}~\bibnamefont
  {\ifmmode~\check{Z}\else \v{Z}\fi{}elezn\'y}}, \bibinfo {author}
  {\bibfnamefont {I.~M.}\ \bibnamefont {Miron}}, \bibinfo {author}
  {\bibfnamefont {T.}~\bibnamefont {Jungwirth}}, \bibinfo {author}
  {\bibfnamefont {J.}~\bibnamefont {Sinova}}, \bibinfo {author} {\bibfnamefont
  {A.}~\bibnamefont {Thiaville}}, \bibinfo {author} {\bibfnamefont
  {K.}~\bibnamefont {Garello}},\ and\ \bibinfo {author} {\bibfnamefont
  {P.}~\bibnamefont {Gambardella}},\ }\bibfield  {title} {\bibinfo {title}
  {Current-induced spin-orbit torques in ferromagnetic and antiferromagnetic
  systems},\ }\href {https://doi.org/10.1103/RevModPhys.91.035004} {\bibfield
  {journal} {\bibinfo  {journal} {Rev. Mod. Phys.}\ }\textbf {\bibinfo {volume}
  {91}},\ \bibinfo {pages} {035004} (\bibinfo {year} {2019})}\BibitemShut
  {NoStop}%
\bibitem [{\citenamefont {Vansteenkiste}\ \emph {et~al.}(2014)\citenamefont
  {Vansteenkiste}, \citenamefont {Leliaert}, \citenamefont {Dvornik},
  \citenamefont {Helsen}, \citenamefont {Garcia-Sanchez},\ and\ \citenamefont
  {Van~Waeyenberge}}]{Vansmumax32014}%
  \BibitemOpen
  \bibfield  {author} {\bibinfo {author} {\bibfnamefont {A.}~\bibnamefont
  {Vansteenkiste}}, \bibinfo {author} {\bibfnamefont {J.}~\bibnamefont
  {Leliaert}}, \bibinfo {author} {\bibfnamefont {M.}~\bibnamefont {Dvornik}},
  \bibinfo {author} {\bibfnamefont {M.}~\bibnamefont {Helsen}}, \bibinfo
  {author} {\bibfnamefont {F.}~\bibnamefont {Garcia-Sanchez}},\ and\ \bibinfo
  {author} {\bibfnamefont {B.}~\bibnamefont {Van~Waeyenberge}},\ }\bibfield
  {title} {\bibinfo {title} {The design and verification of mumax3},\ }\href
  {https://doi.org/10.1063/1.4899186} {\bibfield  {journal} {\bibinfo
  {journal} {AIP Adv.}\ }\textbf {\bibinfo {volume} {4}},\ \bibinfo {pages}
  {107133} (\bibinfo {year} {2014})}\BibitemShut {NoStop}%
\bibitem [{\citenamefont {Ohkoshi}\ \emph {et~al.}(1976)\citenamefont
  {Ohkoshi}, \citenamefont {Kobayshi}, \citenamefont {Katayama}, \citenamefont
  {Hirano}, \citenamefont {Katayama}, \citenamefont {Hirano},\ and\
  \citenamefont {Tsushima}}]{Ohkoshispin1976}%
  \BibitemOpen
  \bibfield  {author} {\bibinfo {author} {\bibfnamefont {M.}~\bibnamefont
  {Ohkoshi}}, \bibinfo {author} {\bibfnamefont {H.}~\bibnamefont {Kobayshi}},
  \bibinfo {author} {\bibfnamefont {T.}~\bibnamefont {Katayama}}, \bibinfo
  {author} {\bibfnamefont {M.}~\bibnamefont {Hirano}}, \bibinfo {author}
  {\bibfnamefont {T.}~\bibnamefont {Katayama}}, \bibinfo {author}
  {\bibfnamefont {M.}~\bibnamefont {Hirano}},\ and\ \bibinfo {author}
  {\bibfnamefont {T.}~\bibnamefont {Tsushima}},\ }\bibfield  {title} {\bibinfo
  {title} {Spin reorientation in {NdCo}5 single crystals},\ }\href
  {https://doi.org/10.1063/1.30484} {\bibfield  {journal} {\bibinfo  {journal}
  {AIP Conf. Proc.}\ }\textbf {\bibinfo {volume} {29}},\ \bibinfo {pages} {616}
  (\bibinfo {year} {1976})}\BibitemShut {NoStop}%
\bibitem [{\citenamefont {Seifert}\ \emph {et~al.}(2013)\citenamefont
  {Seifert}, \citenamefont {Schultz}, \citenamefont {Sch{\"a}fer},
  \citenamefont {Neu}, \citenamefont {Hankemeier}, \citenamefont {R{\"o}ssler},
  \citenamefont {Fr{\"o}mter},\ and\ \citenamefont
  {Oepen}}]{Seifertdomain2013}%
  \BibitemOpen
  \bibfield  {author} {\bibinfo {author} {\bibfnamefont {M.}~\bibnamefont
  {Seifert}}, \bibinfo {author} {\bibfnamefont {L.}~\bibnamefont {Schultz}},
  \bibinfo {author} {\bibfnamefont {R.}~\bibnamefont {Sch{\"a}fer}}, \bibinfo
  {author} {\bibfnamefont {V.}~\bibnamefont {Neu}}, \bibinfo {author}
  {\bibfnamefont {S.}~\bibnamefont {Hankemeier}}, \bibinfo {author}
  {\bibfnamefont {S.}~\bibnamefont {R{\"o}ssler}}, \bibinfo {author}
  {\bibfnamefont {R.}~\bibnamefont {Fr{\"o}mter}},\ and\ \bibinfo {author}
  {\bibfnamefont {H.~P.}\ \bibnamefont {Oepen}},\ }\bibfield  {title} {\bibinfo
  {title} {Domain evolution during the spin-reorientation transition in
  epitaxial {NdCo}5thin films},\ }\href
  {https://doi.org/10.1088/1367-2630/15/1/013019} {\bibfield  {journal}
  {\bibinfo  {journal} {New J. Phys}\ }\textbf {\bibinfo {volume} {15}},\
  \bibinfo {pages} {013019} (\bibinfo {year} {2013})}\BibitemShut {NoStop}%
\bibitem [{\citenamefont {Seifert}\ \emph {et~al.}(2017)\citenamefont
  {Seifert}, \citenamefont {Schultz}, \citenamefont {Sch{\"a}fer},
  \citenamefont {Hankemeier}, \citenamefont {Fr{\"o}mter}, \citenamefont
  {Oepen},\ and\ \citenamefont {Neu}}]{Mariettamicro2017}%
  \BibitemOpen
  \bibfield  {author} {\bibinfo {author} {\bibfnamefont {M.}~\bibnamefont
  {Seifert}}, \bibinfo {author} {\bibfnamefont {L.}~\bibnamefont {Schultz}},
  \bibinfo {author} {\bibfnamefont {R.}~\bibnamefont {Sch{\"a}fer}}, \bibinfo
  {author} {\bibfnamefont {S.}~\bibnamefont {Hankemeier}}, \bibinfo {author}
  {\bibfnamefont {R.}~\bibnamefont {Fr{\"o}mter}}, \bibinfo {author}
  {\bibfnamefont {H.~P.}\ \bibnamefont {Oepen}},\ and\ \bibinfo {author}
  {\bibfnamefont {V.}~\bibnamefont {Neu}},\ }\bibfield  {title} {\bibinfo
  {title} {Micromagnetic investigation of domain and domain wall evolution
  through the spin-reorientation transition of an epitaxial {NdCo}5film},\
  }\href {https://doi.org/10.1088/1367-2630/aa60d5} {\bibfield  {journal}
  {\bibinfo  {journal} {New J. Phys}\ }\textbf {\bibinfo {volume} {19}},\
  \bibinfo {pages} {033002} (\bibinfo {year} {2017})}\BibitemShut {NoStop}%
\bibitem [{\citenamefont {Kumar}\ \emph {et~al.}(2020)\citenamefont {Kumar},
  \citenamefont {Patrick}, \citenamefont {Edwards}, \citenamefont
  {Balakrishnan}, \citenamefont {Lees},\ and\ \citenamefont
  {Staunton}}]{Santoshtorque2020}%
  \BibitemOpen
  \bibfield  {author} {\bibinfo {author} {\bibfnamefont {S.}~\bibnamefont
  {Kumar}}, \bibinfo {author} {\bibfnamefont {C.~E.}\ \bibnamefont {Patrick}},
  \bibinfo {author} {\bibfnamefont {R.~S.}\ \bibnamefont {Edwards}}, \bibinfo
  {author} {\bibfnamefont {G.}~\bibnamefont {Balakrishnan}}, \bibinfo {author}
  {\bibfnamefont {M.~R.}\ \bibnamefont {Lees}},\ and\ \bibinfo {author}
  {\bibfnamefont {J.~B.}\ \bibnamefont {Staunton}},\ }\bibfield  {title}
  {\bibinfo {title} {Torque magnetometry study of the spin reorientation
  transition and temperature-dependent magnetocrystalline anisotropy in
  {NdCo}5},\ }\href {https://doi.org/10.1088/1361-648x/ab7ad6} {\bibfield
  {journal} {\bibinfo  {journal} {J. Phys.: Condens. Matter}\ }\textbf
  {\bibinfo {volume} {32}},\ \bibinfo {pages} {255802} (\bibinfo {year}
  {2020})}\BibitemShut {NoStop}%
\bibitem [{\citenamefont {Timopheev}\ \emph {et~al.}(2016)\citenamefont
  {Timopheev}, \citenamefont {Sousa}, \citenamefont {Chshiev}, \citenamefont
  {Nguyen},\ and\ \citenamefont {Dieny}}]{TimopheevSR2016}%
  \BibitemOpen
  \bibfield  {author} {\bibinfo {author} {\bibfnamefont {A.~A.}\ \bibnamefont
  {Timopheev}}, \bibinfo {author} {\bibfnamefont {R.}~\bibnamefont {Sousa}},
  \bibinfo {author} {\bibfnamefont {M.}~\bibnamefont {Chshiev}}, \bibinfo
  {author} {\bibfnamefont {H.~T.}\ \bibnamefont {Nguyen}},\ and\ \bibinfo
  {author} {\bibfnamefont {B.}~\bibnamefont {Dieny}},\ }\bibfield  {title}
  {\bibinfo {title} {Second order anisotropy contribution in perpendicular
  magnetic tunnel junctions},\ }\href {http://dx.doi.org/10.1038/srep26877}
  {\bibfield  {journal} {\bibinfo  {journal} {Sci. Rep.}\ }\textbf {\bibinfo
  {volume} {6}} (\bibinfo {year} {2016})}\BibitemShut {NoStop}%
\bibitem [{\citenamefont {Coey}(2009)}]{Coey}%
  \BibitemOpen
  \bibfield  {author} {\bibinfo {author} {\bibfnamefont {J.~M.~D.}\
  \bibnamefont {Coey}},\ }\href@noop {} {\emph {\bibinfo {title} {Magnetism and
  Magnetic Materials}}}\ (\bibinfo  {publisher} {Cambridge University Press,
  Cambridge, England},\ \bibinfo {year} {2009})\BibitemShut {NoStop}%
\bibitem [{\citenamefont {Wieser}\ \emph {et~al.}(2010)\citenamefont {Wieser},
  \citenamefont {Vedmedenko},\ and\ \citenamefont
  {Wiesendanger}}]{WieserPRB2010}%
  \BibitemOpen
  \bibfield  {author} {\bibinfo {author} {\bibfnamefont {R.}~\bibnamefont
  {Wieser}}, \bibinfo {author} {\bibfnamefont {E.~Y.}\ \bibnamefont
  {Vedmedenko}},\ and\ \bibinfo {author} {\bibfnamefont {R.}~\bibnamefont
  {Wiesendanger}},\ }\bibfield  {title} {\bibinfo {title} {Domain wall motion
  damped by the emission of spin waves},\ }\href
  {https://doi.org/10.1103/PhysRevB.81.024405} {\bibfield  {journal} {\bibinfo
  {journal} {Phys. Rev. B}\ }\textbf {\bibinfo {volume} {81}},\ \bibinfo
  {pages} {024405} (\bibinfo {year} {2010})}\BibitemShut {NoStop}%
\bibitem [{\citenamefont {Yan}\ and\ \citenamefont {Bauer}(2012)}]{YanPRL2012}%
  \BibitemOpen
  \bibfield  {author} {\bibinfo {author} {\bibfnamefont {P.}~\bibnamefont
  {Yan}}\ and\ \bibinfo {author} {\bibfnamefont {G.~E.~W.}\ \bibnamefont
  {Bauer}},\ }\bibfield  {title} {\bibinfo {title} {Magnonic domain wall heat
  conductance in ferromagnetic wires},\ }\href
  {https://doi.org/10.1103/PhysRevLett.109.087202} {\bibfield  {journal}
  {\bibinfo  {journal} {Phys. Rev. Lett.}\ }\textbf {\bibinfo {volume} {109}},\
  \bibinfo {pages} {087202} (\bibinfo {year} {2012})}\BibitemShut {NoStop}%
\bibitem [{\citenamefont {Gross}(1961)}]{GrossINC1961}%
  \BibitemOpen
  \bibfield  {author} {\bibinfo {author} {\bibfnamefont {E.~P.}\ \bibnamefont
  {Gross}},\ }\bibfield  {title} {\bibinfo {title} {Structure of a quantized
  vortex in boson systems},\ }\href@noop {} {\bibfield  {journal} {\bibinfo
  {journal} {Il Nuovo Cimento}\ }\textbf {\bibinfo {volume} {20}},\ \bibinfo
  {pages} {454} (\bibinfo {year} {1961})}\BibitemShut {NoStop}%
\bibitem [{\citenamefont {Pitaevskii}(1961)}]{PitaevskiiJETP1961}%
  \BibitemOpen
  \bibfield  {author} {\bibinfo {author} {\bibfnamefont {L.~P.}\ \bibnamefont
  {Pitaevskii}},\ }\bibfield  {title} {\bibinfo {title} {Vortex lines in an
  imperfect bose gas},\ }\href@noop {} {\bibfield  {journal} {\bibinfo
  {journal} {Sov. Phys. JETP.}\ }\textbf {\bibinfo {volume} {13}},\ \bibinfo
  {pages} {451} (\bibinfo {year} {1961})}\BibitemShut {NoStop}%
\bibitem [{\citenamefont {Bender}\ \emph {et~al.}(2014)\citenamefont {Bender},
  \citenamefont {Duine}, \citenamefont {Brataas},\ and\ \citenamefont
  {Tserkovnyak}}]{Benderdynamic2014}%
  \BibitemOpen
  \bibfield  {author} {\bibinfo {author} {\bibfnamefont {S.~A.}\ \bibnamefont
  {Bender}}, \bibinfo {author} {\bibfnamefont {R.~A.}\ \bibnamefont {Duine}},
  \bibinfo {author} {\bibfnamefont {A.}~\bibnamefont {Brataas}},\ and\ \bibinfo
  {author} {\bibfnamefont {Y.}~\bibnamefont {Tserkovnyak}},\ }\bibfield
  {title} {\bibinfo {title} {Dynamic phase diagram of dc-pumped magnon
  condensates},\ }\href {https://doi.org/10.1103/PhysRevB.90.094409} {\bibfield
   {journal} {\bibinfo  {journal} {Phys. Rev. B}\ }\textbf {\bibinfo {volume}
  {90}},\ \bibinfo {pages} {094409} (\bibinfo {year} {2014})}\BibitemShut
  {NoStop}%
\bibitem [{\citenamefont {Brataas}\ \emph {et~al.}(2000)\citenamefont
  {Brataas}, \citenamefont {Nazarov},\ and\ \citenamefont
  {Bauer}}]{Brataasfinite2000}%
  \BibitemOpen
  \bibfield  {author} {\bibinfo {author} {\bibfnamefont {A.}~\bibnamefont
  {Brataas}}, \bibinfo {author} {\bibfnamefont {Y.~V.}\ \bibnamefont
  {Nazarov}},\ and\ \bibinfo {author} {\bibfnamefont {G.~E.~W.}\ \bibnamefont
  {Bauer}},\ }\bibfield  {title} {\bibinfo {title} {Finite-element theory of
  transport in ferromagnet--normal metal systems},\ }\href
  {https://doi.org/10.1103/PhysRevLett.84.2481} {\bibfield  {journal} {\bibinfo
   {journal} {Phys. Rev. Lett.}\ }\textbf {\bibinfo {volume} {84}},\ \bibinfo
  {pages} {2481} (\bibinfo {year} {2000})}\BibitemShut {NoStop}%
\bibitem [{Note1()}]{Note1}%
  \BibitemOpen
  \bibinfo {note} {The modified value of the polar angle $\theta $ can be
  calculated by solving Eq.~\protect \textup {\hbox {\mathsurround \z@ \protect
  \normalfont (\ignorespaces \ref {fmllg2}\unskip \@@italiccorr )}} and
  Eq.~\protect \textup {\hbox {\mathsurround \z@ \protect \normalfont
  (\ignorespaces \ref {fmphi}\unskip \@@italiccorr )}} self-consistently for
  $\theta $.}\BibitemShut {Stop}%
\end{thebibliography}%

\end{document}